\documentclass[12pt]{iopart}

\pdfoutput=1

\usepackage[english]{babel}
\usepackage{
amssymb,amsfonts,latexsym}
\usepackage{graphicx}
\usepackage{color}
\usepackage{iopams}
\usepackage{subfigure}
\usepackage{afterpage}
\usepackage{blkarray}
\usepackage{cite}
\usepackage{tikz}
\usetikzlibrary{snakes}
\usetikzlibrary{calc}
\usepackage{bm}
\usepackage{braket}

\definecolor{blue(pigment)}{rgb}{0.2, 0.2, 0.6}
\usepackage{mathrsfs}
\usepackage{times}
\usepackage[colorlinks,
citecolor=blue,linkcolor=blue,urlcolor=blue]{hyperref}
\usepackage{scalerel}
\usepackage{tensor}
\usepackage{etoolbox}

\makeatletter
\def\@mkboth#1#2{}
\newlength\appendixwidth
\preto\appendix{\addtocontents{toc}{\protect\patchl@section}}
\newcommand{\patchl@section}{%
  \settowidth{\appendixwidth}{\textbf{Appendix }}%
  \addtolength{\appendixwidth}{1.5em}%
  \patchcmd{\l@section}{1.5em}{\appendixwidth}{}{\ddt}%
}
\makeatother

\def\eqref#1{(\ref{#1})}

\usepackage{cancel}

\newcommand{\be}{\begin{equation}}
\newcommand{\ee}{\end{equation}}
\newcommand{\bea}{\begin{eqnarray}}
\newcommand{\eea}{\end{eqnarray}}

\newcommand\reallywidehat[1]{\arraycolsep=0pt\relax%
\begin{array}{c}
\stretchto{
  \scaleto{
    \scalerel*[\widthof{\ensuremath{#1}}]{\kern-.5pt\bigwedge\kern-.5pt}
    {\rule[-\textheight/2]{1ex}{\textheight}} 
  }{\textheight} %
}{0.5ex}\\           
#1\\                 
\rule{-1ex}{0ex}
\end{array}
}

\begin{document}

\title[Integrable quenches in nested spin chains I]{Integrable quenches in nested spin chains I:\\ 
	the exact steady states}
\author{Lorenzo Piroli$^{1,2}$, Eric Vernier$^{3}$, Pasquale Calabrese$^{1,4}$, Bal\'{a}zs Pozsgay$^{5,6}$}
\address{$^1$ SISSA and INFN, via Bonomea 265, 34136 Trieste, Italy}
\address{$^2$ Max-Planck-Institut f\"ur Quantenoptik, Hans-Kopfermann-Str. 1, 85748 Garching, Germany}
\address{$^3$ The Rudolf Peierls Centre for Theoretical Physics, Oxford University, Oxford, OX1 3NP, United Kingdom.}
\address{$^4$ International Centre for Theoretical Physics (ICTP), I-34151, Trieste, Italy}
\address{$^5$ Department of Theoretical Physics, Budapest University
	of Technology and Economics, 1111 Budapest, Budafoki \'{u}t 8, Hungary}
\address{$^6$ BME Statistical Field Theory Research Group, Institute of Physics,
	Budapest University of Technology and Economics, H-1111 Budapest, Hungary}
\date{\today}

\begin{abstract}
We consider quantum quenches in the integrable $SU(3)$-invariant spin chain (Lai-Sutherland model) which admits a Bethe ansatz description in terms of two different quasiparticle species, providing a prototypical example of a model solvable by nested Bethe ansatz. We identify infinite families of integrable initial states for which analytic results can be obtained. We show that they include special families of two-site product states which can be related to integrable ``soliton non-preserving'' boundary conditions in an appropriate rotated channel. We present a complete analytical result for the quasiparticle rapidity distribution functions corresponding to the stationary state reached at large times after the quench from the integrable initial states. Our results are obtained within a Quantum Transfer Matrix (QTM) approach, which does not rely on the knowledge of the quasilocal conservation laws or of the overlaps between the initial states and the eigenstates of the Hamiltonian. Furthermore, based on an analogy with previous works, we conjecture analytic expressions for such overlaps: this allows us to employ the Quench Action method to derive a set of integral equations characterizing the quasi-particle distribution functions of the post-quench steady state. We verify that the solution to the latter coincides with our analytic result found using the QTM approach. Finally, we present a direct physical application of our results by providing predictions for the propagation of entanglement after the quench from such integrable states.
\end{abstract}


\maketitle

{\hypersetup{linkcolor=blue(pigment)}
\tableofcontents
}

\section{Introduction}
\label{sec:intro1}
In the past few years, the theory of quantum quenches in isolated integrable models has witnessed unexpected developments \cite{CaEM16}. In fact, many of the results that have been obtained in the most recent literature could hardly be foreseen, say, ten years ago. An important achievement has been the development of analytic approaches to provide explicit predictions for several quantities of interest in the presence of genuine interactions -- a fact that is quite remarkable, given the overwhelming complexity of many-body physics out of equilibrium.
All the theoretical advances have been paralleled by accurate cold-atom experiments accessing the non-equilibrium dynamics of nearly integrable 
models \cite{kinoshita-2006,langen-15,langen-rev,bsjs-18,sbdd-19}.

The first step of this recent revolution has been the introduction of the Quench Action method (QAM) \cite{CaEs13,Caux16}. One of its main achievements has been to show the existence of
a representative eigenstate of the Hamiltonian which effectively captures, in the
thermodynamic limit, the local properties of the system at
long times after a quench. This eigenstate is typically described in terms of the corresponding quasimomentum (or rapidity) distribution function of the stable quasiparticles. In several models with an elementary (i.e. non-nested) Bethe ansatz, it was proved that this description is equivalent to the one in terms of the generalized Gibbs ensemble (GGE) \cite{EsFa16,ViRi16,RDYO07}. The latter is an appropriate statistical ensemble which is analogous to the Gibbs density matrix, but it is built out of all the local and quasilocal conserved operators (or charges) of the Hamiltonian \cite{IDWC15,IMPZ16,IQNB16,IlQC17,PoVW17}. The emerging picture is in fact very similar to what we do in standard thermodynamics, where the thermal Gibbs ensemble for a free gas is represented in terms of the momentum distribution function of its particles~\cite{takahashi-99}.

The merits of the Quench Action approach have been many, such as the study of several instances where the steady state reached at large times after a quench exhibits exotic, non-thermal features \cite{DWBC14,DWBC14_II,PMWK14,BePC16,ac-15,PiCE16,Bucc16,MBPC17,BeTC17}, or even to tackle, in some special cases, the full post-quench dynamics \cite{BeSE14,DeCa14,DePC15,PiCa17}. However, this method is intrinsically limited to those initial states for which the overlaps with the eigenstates of the Hamiltonian can be computed, a task which is in general very hard and has been carried out only in a few cases~\cite{KoPo12,Pozs14,cd-14,PiCa14,Broc14,HoST16,BrSt17,LeKZ15,LeKM16,LeKL18,Pozs18}.

More recently, a quite different technique, termed string-charge duality \cite{IQNB16}, has been devised to directly obtain the post-quench quasiparticle rapidity distribution functions. Within this approach, the post-quench steady state is uniquely fixed by the constraints resulting from a complete set of conservation laws, and has been proven to be applicable to very large classes of initial states \cite{IQNB16,PoVW17,PiVC16,PVCR17}, enlarging considerably the space of physical features that can be captured by analytical inspection. Still, an essential ingredient for this method is the knowledge of a complete set of local and quasilocal conserved charges. While these are completely known in the prototypical case of XXZ Heisenberg spin-$s$ Hamiltonians \cite{Pros11,PrIl13,Pros14,PPSA14,PiVe16,IlMP15,DeCD17} (both in massive and massless regimes), the structure of conservation laws is model-dependent, and in many cases is still not completely understood.

An important class of models which has proven to be hard to tackle by both the QAM and the string-charge duality is provided by nested integrable systems \cite{efgk-05}. These models are interesting from the physical point of view as they present different quasiparticle species, and peculiar features are expected to arise as a result of the interplay between the corresponding bound states. Clearly interesting examples from the experimental point of view are multi-component Fermi and Bose ultra-cold gases \cite{BlDZ08,guan2013,PMCL14}, which provide a largely unexplored arena for realizing interesting non-equilibrium states of matter. 

The prototypical example of a nested system is the $SU(3)$-invariant Lai-Sutherland spin chain \cite{lai-74,sutherland-75}. Even in this simplest case, non-trivial challenges arise for the study of the quench dynamics: for instance, it is worth to stress that until very recently analytical results were restricted to the study of a \emph{single} initial state \cite{MBPC17}. Indeed, an explicit formula for the overlaps between the eigenstates of the Hamiltonian and a particular matrix product state \cite{PVWC06} was conjectured in \cite{LeKM16}, in the different context of the AdS/CFT correspondence (later, this result was extended in \cite{LeKL18} for an infinite discrete family of matrix product states with increasing bond dimension). In turn, this straightforwardly made it possible to apply the QAM in \cite{MBPC17}, and to analyze the corresponding quench problem. While Ref. \cite{MBPC17} showed that the QAM can be applied to nested systems with no additional difficulty once the overlaps are known, it was still limited to one single state.

Paralleling these developments, a third approach to the quench dynamics was introduced in \cite{PiPV17} (see also the previous work \cite{Pozs13}) based on the Quantum Transfer Matrix (QTM) formalism \cite{Klum92,Klum04}. Within this approach, it was possible to define in \cite{PiPV17_II} a class of \emph{integrable initial states} for quench problems, for which analytic results could be obtained. The importance of the definition given in \cite{PiPV17_II}, inspired by well-known field theoretical constructions \cite{GhZa94} (see also \cite{Delf14,Schu15}), lies mainly in the possibility of applying to these states a series of analytic tools to obtain straightforwardly several physical quantities, such as the  Loschmidt echo \cite{PiPV18}. Furthermore, it was shown in \cite{PiPV17} that the QTM approach could be used to analytically extract, with a little effort, the rapidity distribution functions associated to the GGE. For those states for which the QAM and the string-charge duality could be applied, it was shown that the three approaches provide the same result.

In this work we apply the QTM approach to the prototypical $SU(3)$-invariant spin chain \cite{lai-74,sutherland-75}. We derive analytical results for an infinite class of integrable initial states, which can be related to integrable ``soliton non-preserving'' boundary conditions in an appropriate rotated channel.  In particular, based on a few assumptions, we are able to derive the rapidity distribution functions of the post-quench steady state.  

In the light of these analytic results, and based on the experience of previous cases studied in the literature, it is natural to expect that overlap formulas between integrable states and the eigenstates of the Hamiltonian exist also in the nested model of interest in this work. Motivated by this, we conjecture such formulas, and test them numerically for finite system sizes against exact diagonalization calculations. Assuming the validity of our conjecture, we apply the Quench Action method and derive a set of integral equations that characterize the post-quench steady state. As an important consistency check, we show that the solution to the latter coincides with our analytic result found using the QTM approach.

This work is the first of two papers where our results are presented and derived. In this manuscript we focus on the determination of the integrable states and their rapidity distribution functions. In order to do so, we employ certain fusion relations of ``soliton non-preserving'' boundary transfer matrices; the derivation of the latter is non-trivial and will be reported in the second paper \cite{InPrep}, where we focus on the more mathematical aspects of the geometrical structures involved. The fusion relations are in fact also interesting per se, as they allow one to compute the spectrum of the boundary transfer matrices, and are thus relevant also for the equilibrium physics of open integrable Hamiltonians.

The organization of this work is as follows. In Sec.~\ref{sec:intro} we introduce the $SU(3)$-invariant spin chain, and briefly review its Bethe ansatz solution. In Sec.~\ref{sec:integrable_states} we present our construction for the integrable initial states related to soliton-non-preserving boundary conditions in the transverse direction. Next, the main results of our work are consigned to Sec.~\ref{sec:distribution_functions}, where the analytic results for the post-quench rapidity distribution functions are written down explicitly, and analytical checks are provided. In Sec.~\ref{sec:overlaps} we present our conjecture for the overlaps of the integrable states, and carry out the ensuing Quench Action treatment. In Sec. \ref{Sec:entanglement}, as an illustration of a physical application of our results, we present a prediction for the evolution of the entanglement -- a prediction which can  be directly tested numerically. We conclude the paper in Sec. \ref{concl} and relegate some technical details to the appendices.

\section{The model}\label{sec:intro}

\subsection{The Hamiltonian and the nested Bethe ansatz}
We consider the spin-$1$ Lai-Sutherland model \cite{lai-74,sutherland-75}, described by the Hamiltonian  
\be
H_L=\sum_{j=1}^{L}\left[{\bf s}_j\cdot {\bf s}_{j+1}+\left({\bf s}_j\cdot {\bf s}_{j+1}\right)^2\right]-2L\,,
\label{eq:hamiltonian}
\ee
which acts on the Hilbert space $\mathcal{H}_L=h_1\otimes \ldots\otimes h_L$. Here $h_j\simeq \mathbb{C}^3$ is the local (physical) Hilbert space associated with site $j$. The spin-$1$ operators $s^{a}_j$ are given by the standard three-dimensional representation of the $SU(2)$ generators, explicitly
\be
 \fl s^x=\frac{1}{\sqrt{2}}\left(\begin{array}{c c c}0&1&0\\1&0&1\\0&1&0\end{array}\right),\quad s^y=\frac{1}{\sqrt{2}}\left(\begin{array}{c c c}0&-i&0\\i&0&-i\\0&i&0\end{array}\right),\quad s^z=\left(\begin{array}{c c c}1&0&0\\0&0&0\\0&0&-1\end{array}\right)\,.
\label{eq:spin_op}
\ee
In the following, we also define 
\be
\ket{1} =\left(\begin{array}{c}1\\0\\0\end{array}\right),\quad \ket{2} =\left(\begin{array}{c}0\\1\\0\end{array}\right), \quad \ket{3} =\left(\begin{array}{c}0\\0\\1\end{array}\right)\,,
\ee
and
\be
|\alpha_1\,,\alpha_2\,,\ldots \alpha_L\rangle=|\alpha_1\,\rangle\otimes |\alpha_2\rangle\otimes \ldots \otimes |\alpha_L\rangle\,.
\label{eq:notation_2}
\ee
The Hamiltonian \eqref{eq:hamiltonian} is invariant under global action of $SU(3)$, as it is manifest from the equivalent expression
\be
H_L=-L+\sum_{j=1}^{L}\mathcal{P}_{j,j+1}\,,
\label{eq:hamiltonian_permutation}
\ee
where $\mathcal{P}_{j,j+1}$ are permutation operators
\be
\mathcal{P}_{j,j+1}|a\rangle_j\otimes |b\rangle_{j+1}=|b\rangle_j\otimes |a\rangle_{j+1}\,.
\label{eq:permutation_matrix}
\ee
The Hamiltonian \eqref{eq:hamiltonian} is particularly interesting, as its quasiparticle content consists of different species, providing a prototypical example of a model solvable by “nested” Bethe ansatz~\cite{efgk-05, kr-81,johannesson-86,johannesson2-86}.

To each eigenstate of the Hamiltonian we can associate two sets of complex parameters called rapidities, ${\boldsymbol k}_N=\{k_j\}_{j=1}^N$ and ${\boldsymbol \lambda}_M=\{\lambda_j\}_{j=1}^{M}$. They generalize the concept of quasimomenta for free particles, parametrizing the exact eigenfunctions as
\bea
\fl | \boldsymbol k_N, \boldsymbol \lambda_N\rangle = \sum_{1\leq n_1<\ldots <n_N\leq L}\ \sum_{1\leq m_1<\ldots <m_M\leq N}\ 
\sum_{\mathcal P \in \mathcal S_N} 
\left(\prod_{1\leq r < l \leq N}\frac{k_{\mathcal P (l)}-k_{\mathcal P (r)}-i}{k_{\mathcal P (l)}-k_{\mathcal P (r)}}\right)\nonumber\\
\times \braket{\boldsymbol m|\boldsymbol k_{\mathcal P}, \boldsymbol \lambda} \prod_{r=1}^N\left(\frac{k_{\mathcal{P}(r)}+i/2}{k_{\mathcal{P}(r)}-i/2}\right)^{n_r}\prod_{r=1}^{M}(E^{2}_3)_{n_{m_r}}\prod_{s=1}^{N}(E^{1}_2)_{n_{s}}|\Omega\rangle\,,
\label{eq:bethe_state}
\eea
where we introduced the reference state
\be
|\Omega\rangle=|1,1, \ldots 1\rangle\,,
\label{eq:reference_state}
\ee
and the operators
\be
E^{i}_j\equiv|j\rangle\langle i|\,,
\label{eq:eij_operator}
\ee
together with the functions
\bea
\braket{\boldsymbol m|\boldsymbol k_{\mathcal P}, \boldsymbol \lambda} &=& \sum_{\mathcal R \in \mathcal S_M} A(\boldsymbol \lambda_{\mathcal R})\prod_{\ell=1}^M F_{\boldsymbol k_{\mathcal P}}(\lambda_{\mathcal R(\ell)}; m_\ell)\,,\label{eq:second_sum}\\
F_{\boldsymbol k}(\lambda,s)&=&\frac{-i }{\lambda - k_s - i /2}\prod_{n=1}^{s-1} \frac{\lambda -  k_n + i /2}{\lambda - k_n - i /2}\,,\\
A(\lambda) &=&\prod_{1\leq r < l \leq M} \frac{\lambda_l-\lambda_r- i}{\lambda_l-\lambda_r}\,.
\eea
The sums in \eqref{eq:bethe_state} and \eqref{eq:second_sum} are over the sets $\mathcal{S}_N$, $\mathcal{S}_M$ of permutations $\mathcal{P}$ and $\mathcal{Q}$ of $N$ and $M$ elements respectively. The wave function \eqref{eq:bethe_state} is interpreted as made of $N$ quasiparticles of rapidities $k_j$ and $M$ quasiparticles of rapidities $\lambda_j$. Note that the numbers of the two kinds of rapidities can be restricted to satisfy
\be
N\leq \frac{2}{3}L\,, \qquad M\leq \frac{N}{2}\,,\label{eq:equator_condition}
\ee 
where $L$ is the length of the chain. Indeed, if \eqref{eq:equator_condition} is not satisfied for a given eigenstate, the latter lies in the wrong ``side of the equator'' \cite{Baxt02}, and one can choose a different reference state for which the same eigenstate is represented with a smaller number of quasi-particles.

In analogy with the well-known Bethe ansatz solution of the spin-$1/2$ Heisenberg chain \cite{kbi-93}, the wave-function \eqref{eq:bethe_state} corresponds to an eigenstate of the Hamiltonian if the rapidities satisfy the following (nested) Bethe equations
\bea
\left(\frac{k_{j}+i/2}{k_{j}-i/2}\right)^{L}=\prod_{\scriptstyle p=1\atop \scriptstyle p\ne j}^{N} \frac{k_j- k_p+ i}{k_j- k_p- i} \prod_{\ell = 1}^{M} \frac{\lambda_\ell- k_j+ i/2}{\lambda_\ell- k_j- i/2}\,, \quad j=1,\ldots,N\,,\label{eq:bethe_equations1}\\
1=\prod_{j=1}^N \frac{k_j-\lambda_\ell- i/2}{k_j-\lambda_\ell+ i/2}\prod_{\scriptstyle m=1\atop \scriptstyle m\neq \ell}^M  \frac{\lambda_\ell-\lambda_m- i}{\lambda_\ell-\lambda_m+ i}\,, \quad \ell=1,\ldots,M\,.
\label{eq:bethe_equations2}
\eea
Finally, we note that the observables of a given eigenstate can be expressed in terms of the rapidities. For example, the energy and momentum eigenvalues read
\bea
E&=&-\sum_{j=1}^{N}\frac{1}{k_j^2+1/4}\,,\\
P&=&\left[\sum_{j=1}^{N}i\ln\left[\frac{k_{j}+i/2}{k_{j}-i/2}\right]\right]\qquad (\textrm{mod}\,2\pi)\,.\label{eq:energy_eigenvalue}
\eea

\subsection{The thermodynamic description}

In the thermodynamic limit, defined by $L,N,M\to \infty$ keeping the ratios $D_1=N/L$ and $D_2=M/L$ constant, the number of rapidities associated to a given eigenstate grows to infinity, arranging themselves in the complex plane according to specific patterns. The mathematical framework to study this limit is provided by the thermodynamic Bethe ansatz formalism \cite{takahashi-99}, which we briefly review in the specific case of the $SU(3)$-invariant model, referring the reader to Refs.~\cite{johannesson-86,johannesson2-86, afl-83,jls-89,mntt-93,dn-98} for more detail.

First, we recall that for large $L$, both sets of rapidities form patterns called strings, corresponding to bound states of quasiparticles. Each eigenstate will be in general formed by $M_n^{(1)}$ bound-states of quasiparticles (strings) of the first species and $M_n^{(2)}$ strings of the second one. In turn, each string will contain a number $n$ of rapidities, parametrized as
\bea
\fl k^{n,\ell}_{\alpha}=k^{n}_{\alpha}+i\left(\frac{n+1}{2}-\ell\right)+\delta_{1,\alpha}^{n,\ell}\,,\quad \ell=1,\ldots\, n\,,\qquad \alpha=1,\ldots,M_{n}^{(1)}\,,\label{eq:string_1}\\
\fl \lambda^{n,\ell}_{\alpha}=\lambda^{n}_{\alpha}+i\left(\frac{n+1}{2}-\ell\right)+\delta_{2,\alpha}^{n,\ell}\,,\quad \ell=1,\ldots\, n\,,\qquad \alpha=1,\ldots,M_{n}^{(2)}\,.\label{eq:string_2}
\eea
Here the real numbers $k^{n}_{\alpha}$, $\lambda^{n}_{\alpha}$ are the string centers, which can be interpreted as the quasimomenta of the bound-states, while $\delta_{r,\alpha}^{n,\ell}$ are deviations from a perfect string. Within the  string hypothesis, one assumes that such deviations can be neglected in the thermodynamic limit, and one can switch to a description based uniquely on the string centers.

In the thermodynamic limit the string centers become continuous variables on the real line, distributed according to rapidity distribution functions $\rho^{(1)}_n(k)$ and $\rho^{(2)}_n(\lambda)$. Analogously to the case of free systems, we also introduce the hole distribution functions $ \rho^{(1)}_{h,n}(k)$ and $ \rho^{(2)}_{h,n}(\lambda)$, which correspond to the distributions of holes, namely values of the rapidity for which there is no particle. In the thermodynamic limit, the Bethe equations \eqref{eq:bethe_equations1}, \eqref{eq:bethe_equations2} are cast into the following linear integral equations for the distributions $ \rho^{(1)}_{n}(k)$ and $ \rho^{(2)}_{n}(\lambda)$ \cite{mntt-93}
\bea
\rho_{t,n}^{(1)}(\lambda)&=&a_n(\lambda)-\sum_{m=1}^{\infty}\left(a_{n,m}\ast\rho_m^{(1)}\right)(\lambda)+\sum_{m=1}^{\infty}\left(b_{n,m}\ast\rho_m^{(2)}\right)(\lambda)\,,\label{eq:TBAexplicit1}\\
\rho_{t,n}^{(2)}(\lambda)&=&-\sum_{m=1}^{\infty}\left(a_{n,m}\ast\rho_m^{(2)}\right)(\lambda)+\sum_{m=1}^{\infty}\left(b_{n,m}\ast\rho_m^{(1)}\right)(\lambda)\,.\label{eq:TBAexplicit2}
\eea
Here we employed the standard definitions
\bea
\rho^{(r)}_{t,n}(k)=\rho^{(r)}_{n}(k)+\rho^{(r)}_{h,n}(k)\,,\quad r=1,2\,,\quad n=1,\ldots , +\infty\,,
\eea
together with
\be
\left(f\ast g\right)(\lambda)=\int_{-\infty}^{\infty}{\rm d}\mu f(\lambda-\mu)g(\mu)\,,
\label{eq:convolution}
\ee
and
\bea
\fl a_{n,m}(\lambda)=(1-\delta_{nm})a_{|n-m|}(\lambda)+2a_{|n-m|+2}(\lambda)+\ldots
+2a_{n+m-2}(\lambda)+a_{n+m}(\lambda)\,,\label{eq:a_mn}\\
\fl b_{n,m}(\lambda)=a_{|n-m|+1}(\lambda)+a_{|n-m|+3}(\lambda)+\ldots+a_{n+m-1}(\lambda)\,,
\label{eq:b_mn}
\eea
where
\bea
a_{n}(\lambda)&=&\frac{1}{2\pi}\frac{n}{\lambda^2+n^2/4}\,.
\label{eq:a_function}
\eea
In the following, it will also be useful to work with the functions
\bea
\eta^{(r)}_n(x) & = &\frac{\rho^{(r)}_{h,n}(x)}{\rho^{(r)}_{n}(x)}\,,\quad r=1,2\,,\quad n=1,2,\ldots ,+\infty\,.
\label{eq:eta_functions}
\eea

As in the well-known case of the spin-$1/2$ Heisenberg chain, the Bethe equations \eqref{eq:TBAexplicit1}, \eqref{eq:TBAexplicit2} can be cast in a partially decoupled form which is more convenient for numerical analysis. The derivation of the partially decoupled equations is standard (see for example \cite{MBPC17}) and will not be reported here. The final result reads
\bea
\rho_{t,n}^{(1)}(\lambda)&=&\delta_{n,1}s(\lambda)+s\ast\left( \rho_{h,n-1}^{(1)}+\rho_{h,n+1}^{(1)}\right)(\lambda)+s\ast \rho_{n}^{(2)}(\lambda)\,,\label{eq:tri-diagonal_BT1}\\
\rho_{t,n}^{(2)}(\lambda)&=&s\ast\left( \rho_{h,n-1}^{(2)}+\rho_{h,n+1}^{(2)}\right)(\lambda)+s\ast \rho_{n}^{(1)}(\lambda)\,,
\label{eq:tri-diagonal_BT2}
\eea
where
\be
s(\lambda)=\frac{1}{2{\cosh}\left(\pi\lambda\right)}\,,
\label{eq:s_function}
\ee
and where the following convention is understood
\be
\rho^{(r)}_{h,0}(\lambda)\equiv 0\,,\qquad\qquad r=1,2\,.
\label{eq:convention_rhot0}
\ee
The rapidity distribution functions are believed to completely characterize the properties of a given macrostate in the thermodynamic limit; more precisely, all local observables can be in principle computed once they are known. For example, the densities of quasiparticles of the two-species, $D_1$ and $D_2$, and the energy density $e$ can be obtained in the thermodynamic limit as
\bea
D_1&=& \lim_{\rm th} \frac{{N}}{L}=\sum_{n=1}^{+\infty}n\int_{-\infty}^{+\infty}\!\!\!\!{\rm d}k\,\, \rho_{n}^{(1)}(k)\,,\label{eq:density1}\\
D_2&=& \lim_{\rm th} \frac{{M}}{L}=\sum_{n=1}^{+\infty}n\int_{-\infty}^{+\infty}\!\!\!\!{\rm d}\lambda\,\, \rho_{n}^{(2)}(\lambda)\,,\label{eq:density2}\\
e&=& \lim_{\rm th} \frac{E}{L}=\sum_{n=1}^{+\infty}\int_{-\infty}^{+\infty}\,{\rm d}k\,\, \rho_{n}^{(1)}(k)\varepsilon_{n}(k)\,.
\label{eq:energy_density}
\eea
where
\be
\varepsilon_{n}(k)=-\frac{n}{k^2+n^2/4}\,.
\label{eq:epsilon_energy}
\ee
\section{The integrable initial states}\label{sec:integrable_states}

One of the main goals of this work is to identify families of initial states for which the quench dynamics can be tackled analytically. 
In the recent Refs. \cite{LeKM16,LeKL18} exact overlap formulas have been derived for the following family of matrix product states
\bea
\ket{\Psi_0} &=& \frac{1}{\sqrt\mathcal N}\,{\rm tr}_0\left[\prod_{j=1}^L\Big(S_0^1\ket{1}_j+S_0^2\ket{2}_j+S_0^3\ket{3}_j\Big)\right]\nonumber\\
&=&\frac{1}{\sqrt\mathcal N} \sum_{\{\alpha_j\}}{\rm tr}_0\left[S_0^{\alpha_1}S_0^{\alpha_2}\ldots S_0^{\alpha_L}\right]\ket{{\alpha_1},{\alpha_2}\ldots {\alpha_L}}\,.
\label{eq:initial_state}
\eea
Here $S_0^{1,2,3}$ are the generators of $SU(2)$ in the irreducible spin-$k$ representation, satisfying $[S^a_0,S^b_0]=i\varepsilon_{abc}S^c_0$,
$a,b,c=1\dots 3$, and acting on the auxiliary space $h_0\simeq \mathbb{C}^{2k+1}$. 
For $k=1/2$ they are proportional to the Pauli matrices as $S^a=\sigma^a/2$, where
\be
\sigma^{1}=\left(\begin{array}{c c}0&1\\1&0\end{array}\right)\,,\quad \sigma^{2}=\left(\begin{array}{c c}0&-i\\i&0\end{array}\right)\,,\quad \sigma^{3}=\left(\begin{array}{c c}1&0\\0&-1\end{array}\right)\,.
\ee
The trace in \eqref{eq:initial_state} is over the auxiliary space
$h_0$, and the $k$- and $L$-dependent normalization
constant $\mathcal{N}$ ensures $\langle\Psi_0|\Psi_0\rangle=1$. 
So far  the quench dynamics from the initial state \eqref{eq:initial_state} has been studied 
only for the simplest case of $k=1/2$ \cite{MBPC17}, because  at that time the overlaps were only known
for this specific state \cite{LeKM16}.

As an important subsequent development, which is the starting point of our investigation, a definition of integrable initial states has been introduced in \cite{PiPV17_II}:
they are identified as those matrix product states with finite bond dimension that are annihilated by all the local charges of the model which are odd under space reflection.  This integrability condition was proven in \cite{LeKL18} for the states \eqref{eq:initial_state} for arbitrary $k$. 
It is therefore a pressing issue to understand how these states can be embedded into the integrability framework.

In the following, we initiate the program of identifying and analyzing all integrable initial states of the $SU(3)$-invariant
Hamiltonian \eqref{eq:hamiltonian}. In this work, we investigate the initial states which are obtained as the product of real-space two-site blocks; a systematic study of the MPSs  \eqref{eq:initial_state} is instead presented in \cite{InPrep_II}, where their relation to integrable boundaries is finally clarified. Two-site product states are simpler than the MPSs above; nevertheless they are experimentally relevant and their study has not been performed yet. 

In Ref.~\cite{PiPV17_II} it has been shown how to obtain integrable two-site states for spin models whose $R$-matrix (cf. Sec.~\ref{sec:qtm_construction}) satisfies a crossing relation of the form
\be
R^{T_1}_{1,2}(u)=\gamma(u) W^{-1}_1R_{1,2}(-u-\eta)W_1\,,
\label{eq:crossin_relation}
\ee
where $W$ is some invertible matrix and $\eta$ some complex parameter. The main idea of \cite{PiPV17_II} is that integrable states can be obtained by starting from integrable boundary conditions in an appropriate transverse direction and \eqref{eq:crossin_relation} was used as a technical condition to realize this construction.
Unfortunately, in the case of $SU(3)$-invariant spin chains, the relation \eqref{eq:crossin_relation} is not satisfied, so that our first goal is to generalize the derivation of \cite{PiPV17_II} without using the condition \eqref{eq:crossin_relation}.  It turns out that a generalization is possible, provided that one considers a different type of ``soliton-non-preserving'' integrable boundary conditions. The latter, together with a brief review of boundary Bethe ansatz for $SU(3)$-invariant spin chains~\cite{Doik00,AACD04,AACD04_2} is reviewed in the following. 

\subsection{The quantum transfer matrix construction}\label{sec:qtm_construction}

In this section we carry out the construction to relate initial states after a quantum quench to
boundary conditions in an appropriate crossed direction, by generalizing to the case of the $SU(3)$-Hamiltonian \eqref{eq:hamiltonian} the analysis of \cite{PiPV17}, \cite{PiPV17_II}.

As a first step, we consider the quantity
\be
\mathcal{Z}(\beta)=\langle \Psi_0|e^{-\beta H_L}|\Psi_0\rangle\,.
\label{eq:euclidean_partition_function}
\ee
In a relativistic field-theoretical setting, following the classical work by Ghoshal and Zamolodchikov \cite{GhZa94}, one could interpret \eqref{eq:euclidean_partition_function} in two ways. First, one could think of \eqref{eq:euclidean_partition_function} as a ``Loschmidt amplitude'' at (in general complex) time $\beta$, where $|\Psi_0\rangle$ plays the role of the initial state and where the dynamics is governed by the bulk Hamiltonian $H_L$. However, due to Lorentz invariance, one could also ``rotate the picture'' and interpret \eqref{eq:euclidean_partition_function} as arising from an Hamiltonian dynamics where the space and time directions are exchanged: using this interpretation the dynamics is driven by an open Hamiltonian, where the specific boundary conditions are determined by the initial state \cite{GhZa94}. Integrable initial states are those for which the latter are integrable.

As it was noted in \cite{PiPV17_II}, contrary to the field-theory framework of \cite{GhZa94}, on the lattice there is an intrinsic asymmetry between time, which is continuous, and space, which is discrete. In order to relate space and time directions, it is thus necessary to discretize the latter; this is done in a standard way by means of the following Suzuki-Trotter decomposition \cite{FuKl99}
\be
e^{-\beta H_L}=\lim_{N\to\infty}\rho_{N,L}\,,
\label{eq:ST_dec}
\ee
where
\be
\rho_{N,L}=\left[\bar{t}\left(-\frac{i\beta}{N}\right)t\left(-\frac{i\beta}{N}\right)\right]^{N/2}\,.
\label{eq:rhoNL}
\ee
Here we introduced some fundamental objects of the algebraic Bethe ansatz \cite{efgk-05}, which are essential for our constructions. In particular, we defined the transfer matrices
\bea
t(\lambda)={\rm tr}_j\left[\widehat{R}_{jL}(\lambda)\ldots \widehat{R}_{j1}(\lambda)\right]\,,\label{eq:periodic_transfer_matrix}\\
\bar{t}(\lambda)={\rm tr}_j\left[\widehat{R}_{1j}(\lambda)\ldots \widehat{R}_{Lj}(\lambda)\right]\label{eq:periodic_bar_transfer_matrix}\,,
\eea
where the trace is taken over the auxiliary space $h_j\simeq \mathbb{C}^3$, while $\widehat{R}_{ij}(\lambda)$ is the $R$-matrix of the $SU(3)$-invariant model
\begin{equation}
\widehat{R}_{12}(\lambda)=\frac{1}{\lambda+i}\left(\lambda + i \mathcal{P}_{12}\right)\,.
\end{equation}
Here $\mathcal{P}_{12}$ is the permutation matrix defined in \eqref{eq:permutation_matrix}. In the following, it will also be useful to define
\be
R_{1,2}(\lambda)=\lambda+i\mathcal{P}_{1,2}\,.
\label{eq:Rmatrix}
\ee

The $R$-matrix $R_{1,2}(\lambda)$ acts on the product of two fundamental representations. For $SU(3)$-invariant spin chains, there exists a different $R$-matrix, which intertwines the product of a fundamental representation and its conjugate \cite{VeWo92,AbRi96,Doik00_f}
\be
\bar{R}_{12}(\lambda):=R_{12}^{t_1}\left(-\lambda-\frac{3i}{2}\right)
\,.
\label{eq:crossing}
\ee
Here we introduced the transposition
\be
A^{t}=V^{-1}A^{T}V\,,
\ee
where $A^{T}$ is the usual transposed of the matrix $A$, while
\be
V=\left(
\begin{array}{ccc}
	0 & 0 & 1 \\
	0 & 1 & 0 \\
	1 & 0 & 0
\end{array}
\right)\,.
\label{eq:v_matrix}
\ee
With this definition, one can straightforwardly rewrite the transfer matrix $\bar{t}(-i\beta/N)$ in \eqref{eq:rhoNL} in terms of the $R$-matrices $\bar{R}_{ij}(\lambda)$ as
\be
\fl \bar{t}(-i\beta/N)=\frac{1}{\left(-i\beta/N+i\right)^L}{\rm tr}_j\left[\bar{R}_{jL}(i\beta/N-i3/2)\ldots \bar{R}_{j1}(i\beta/N-i3/2)\right]\,.
\label{eq:equivalent_boundary}
\ee

\begin{figure}
	\centering
		\begin{tikzpicture}[scale=0.64]

		\foreach \x in {1, 2.5, 4,...,8.5} { 
			\draw (\x,-0.6)  -- (\x,3.5);
		}
	
		\foreach \x in {1,4,7}{
			\draw[rounded corners=2pt] (\x,-0.6) -- ( \x+0.375,-0.85) -- (\x+0.75,-1.1) -- ( \x+1.125,-0.85) -- ( \x+1.5,-0.6);
			\draw[rounded corners=2pt] (\x,3.5) -- ( \x+0.375,3.75) -- (\x+0.75,4) -- ( \x+1.125,3.75) -- ( \x+1.5,3.5);
		}
		\foreach \x in {1,4,7}{
			\node at (\x+0.75, -1.5) {$| \psi_0\rangle$};
			\node at (\x+0.75, 4.4) {$| \psi_0\rangle$};
		}
		\foreach \x in {0,1,2,3}{
			\draw (0.5,\x)  -- (9,\x);
		}

		\foreach \x in {0,1,2,3}{
			\draw[dashed] (-0.1,\x) -- (0.4,\x) ;
			\draw[dashed]  (9.1,\x) -- (9.7,\x) ;
		}
		\draw[<-> , line width =0.8, draw=black] (10.2,-1) -- (10.2,4.);
		\draw[<-> , line width =0.8, draw=black] (0.5,-2.8) -- (9,-2.8);
		\node at (10.7, 1.5) {$N$};
		\node at (4.75, -3.3) {$L$};


		\foreach \x in {16,18,20,22}{
			\draw (\x, -2) -- (\x, 4);
		}

		\foreach \x in {-1.5,-0.5,0.5,...,3.5}{
			\draw (15.5,\x) -- (22.5, \x);
		};
		\foreach \x in {-1.5,0.5,...,2.5}{
			\draw[rounded corners=2pt] (22.5,\x) -- (22.75,\x+0.25)  -- (23,\x+0.5) --  
			(22.75, \x+0.75) -- (22.5,\x+1);
			\node at (23.8,\x+0.5) {$|\psi_0\rangle $};
			\draw[rounded corners=2pt] (15.5,\x) -- (15.25,\x+0.25)  -- (15,\x+0.5) --  
			(15.25, \x+0.75) -- (15.5,\x+1);
			\node at (14.2,\x+0.5) {$\langle \psi_0| $};
		}
		\foreach \x in {16,18,20,22}{
			\draw[dashed] (\x,-2.6) -- (\x,-2);
			\draw[dashed] (\x,4) -- (\x,4.6);
		};
		\draw[<-> , line width =0.8, draw=black] (24.8,-2) -- (24.8,4.);
		\draw[<-> , line width =0.8, draw=black] (15.5,-2.8) -- (22.5,-2.8);
		\node at (19, -3.3) {$N$};
		\node at (25.2, 1) {$L$};
		\end{tikzpicture}
		
		\caption{Pictorial representation of the partition function \eqref{eq:euclidean_partition_function} for $L=6$ and finite Trotter number $N=4$. Dashed lines correspond to spaces that are traced over. On the left \eqref{eq:euclidean_partition_function} is interpreted as a ``Loschmidt amplitude'' where $|\Psi_0\rangle$ plays the role of the initial state and where the dynamics is governed by the bulk Hamiltonian $H_L$. On the right, we report the same partition function observed in the ``rotated direction'': in this picture, the dynamics is  driven by a two-row transfer matrix , where the specific boundary conditions are determined by the initial state.}
		\label{fig:2d_classical}
\end{figure}
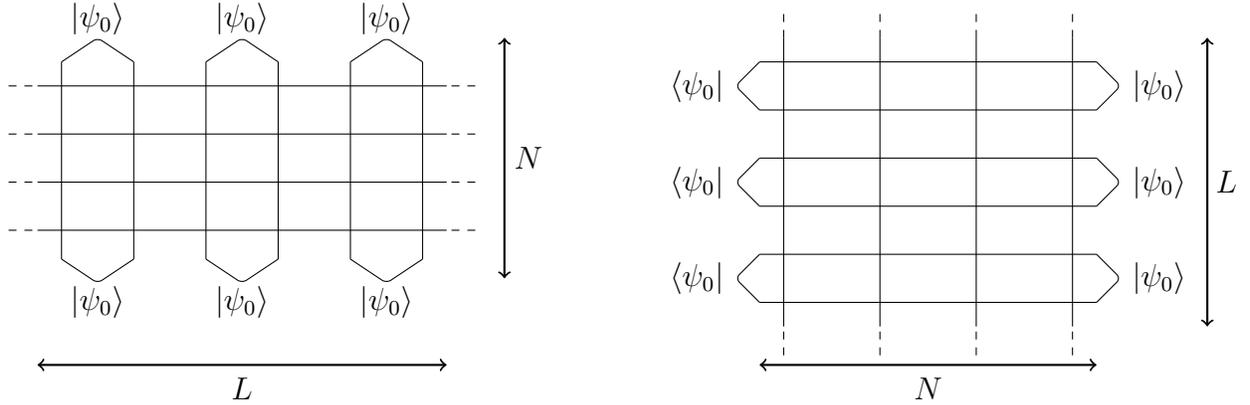

We now focus on two-site product initial states
\be
|\Psi_0\rangle=|\psi\rangle_{1,2}\otimes|\psi\rangle_{3,4}\ldots \otimes |\psi\rangle_{L-1,L}\,.
\label{eq:product_states}
\ee
Replacing $e^{-\beta H_L}$ in \eqref{eq:euclidean_partition_function} with a Trotterized approximation $\rho_{N,L}$ (for finite-$N$), one can see that \eqref{eq:euclidean_partition_function} becomes the partition function of a two-dimensional lattice, where the role of time and space can be exchanged, as it is displayed in Fig.~\ref{fig:2d_classical}. Going further, exploiting \eqref{eq:equivalent_boundary},  it is straightforward to derive the relation
\bea
\langle\Psi_0|\rho_{N,L}|\Psi_0\rangle&=&\langle \Psi_0|\left[\bar{t}\left(-\frac{i\beta}{N}\right)t\left(-\frac{i\beta}{N}\right)\right]^{N/2}|\Psi_0 \rangle
={\rm tr}\left[\mathcal{T}^{L/2}\right] \,,
\label{eq:rhoNLQTM}
\eea 
where
\be
\mathcal{T}=\frac{\langle\psi_0|T^{\rm QTM}_1(0)\otimes T^{\rm QTM}_{2}(0)|\psi_0\rangle}{(1-\beta/N)^{2N}}\,,
\label{eq:mathcalT}
\ee
and
\be 
T^{\rm QTM}_j(\lambda)=R_{N,j}(\lambda-\mu)\bar{R}_{N-1,j}(\lambda-\bar{\mu})\ldots R_{2j}(\lambda-\mu)\bar{R}_{1j}(\lambda-\bar{\mu})\,.
\label{eq:q_monodromy}
\ee 
Here we introduced the inhomogeneities
\bea
\mu&=&-i\beta/N\\
\bar{\mu}&=&i\beta/N-3i/2\,.
\eea

Eq. ~ \eqref{eq:rhoNLQTM} encodes the fact that, in the rotated picture, the evolution can be thought of as generated by a two-row transfer matrix $\mathcal{T}$. Following \cite{PiPV17_II}, it is thus natural to wonder whether there exist states such that \eqref{eq:mathcalT} is equal to an integrable boundary transfer matrix. It turns out that this is the case, and that for such states $\mathcal{T}$ becomes equivalent to a particular kind of transfer matrix, called \emph{soliton non-preserving} \cite{Doik00,AACD04}. This is explained in the following sections, where we first introduce the necessary technical ingredients needed for our analysis.

\subsection{The boundary algebraic Bethe ansatz}

In the previous section, we have already introduced the periodic transfer matrix $t(\lambda)$; its relevance is due to the fact that it generates the Hamiltonian \eqref{eq:hamiltonian} through the well-known relation~\cite{FuKl99}
\be
H=i\frac{\rm d}{{\rm d}\lambda}\log t(\lambda)\big|_{\lambda=0}\,.
\ee
In general, it is possible to define an analogous \emph{boundary} transfer matrix which generates an Hamiltonian with the same bulk terms, but instead of being translationally invariant displays some open boundary conditions. The framework to study such integrable Hamiltonians is called the boundary algebraic Bethe ansatz, and is reviewed in this section.

In the case of $SU(\mathcal{N})$-invariant spin chains (with $\mathcal{N}\geq 3$), the technical tools relevant for our analysis were developed in \cite{Doik00,AACD04,AACD04_2}.  As a main difference with respect to the case of $SU(2)$-invariant spin chains, there exist two inequivalent boundary transfer matrices, which correspond to two kinds of open boundary conditions: these are called ``soliton-preserving'' and ``soliton-non-preserving'' respectively \cite{AACD04}. For completeness, both of them are briefly reviewed here, while we will see that only the latter will be relevant for our purposes. In the following, we only report the main formulas that are directly connected to our work: the interested reader is referred to the literature for a more comprehensive treatment \cite{Doik00,AACD04,AACD04_2}.

\paragraph{Soliton-preserving boundary conditions.} 

We begin by discussing the case of soliton-preserving boundary conditions  \cite{dn-98}. These correspond to the standard case initially introduced for spin-$1/2$ Heisenberg chains in \cite{Skly88}. Given a system of $N$ sites,  one defines the soliton-preserving  boundary transfer matrices as
\bea
\tau_{\rm SP}(\lambda)&=&{\rm tr}_{a}\left\{K_a^{+}(\lambda)T_a(\lambda)K_a^{-}(\lambda)\hat{T}_{a}(\lambda)\right\}\,,\label{eq:det_sp_bc_transfer}\\
\bar{\tau}_{\rm SP}(\lambda)&=&{\rm tr}_a\left\{K_{\bar{a}}^{+}(\lambda)T_{\bar{a}}(\lambda)K_{\bar{a}}^{-}(\lambda)\hat{T}_{\bar{a}}(\lambda)\right\}\,,
\label{eq:det_sp_bc_transfer_bar}
\eea
where
\bea
 T_a(\lambda)=R_{aN}(\lambda-\xi_{N})R_{a(N-1)}(\lambda-\xi_{N-1})\ldots R_{a2}(\lambda-\xi_2) R_{a1}(\lambda-\xi_1)\,,\\
 \hat{T}_{a}(\lambda)=R_{1a}(\lambda+\xi_1)R_{2a}(\lambda+\xi_2)\ldots R_{(N-1)a}(\lambda+\xi_{N-1}) R_{Na}(\lambda+\xi_{N})\,,\\
 T_{\bar{a}}(\lambda)=\bar{R}_{aN}(\lambda-\xi_{N})\bar{R}_{a(N-1)}(\lambda-\xi_{N-1})\ldots \bar{R}_{a2}(\lambda-\xi_2) \bar{R}_{a1}(\lambda-\xi_1)\,,\\
 \hat{T}_{\bar{a}}(\lambda)=\bar{R}_{1a}(\lambda+\xi_1)\bar{R}_{2a}(\lambda+\xi_2)\ldots \bar{R}_{(N-1)a}(\lambda+\xi_{N-1}) \bar{R}_{Na}(\lambda+\xi_{N})\,.
\eea
Here we introduced the inhomogeneities $\xi_j$ which can be thought of as free variables parametrizing the boundary transfer matrices.  Furthermore, the trace in \eqref{eq:det_sp_bc_transfer} and \eqref{eq:det_sp_bc_transfer_bar} is taken over the auxiliary space $h_a\simeq \mathbb{C}^3$, while $K_a^{\pm}(\lambda)$, $K_{\bar{a}}^{\pm}(\lambda)$ are $3\times 3$ matrices.

Importantly, the $K$-matrices $K_a^{\pm}(\lambda)$, $K_{\bar{a}}^{\pm}(\lambda)$ have to be chosen in such a way that transfer matrices with different spectral parameter commute. This can be done by setting $K_{a}^-(\lambda)=K_{a}(\lambda)$, $K_{\bar{a}}^-(\lambda)=K_{\bar{a}}(\lambda)$,  $K_{a}^+(\lambda)=K^t_{a}(-\lambda-i3/2)$ and $K_{\bar{a}}^+(\lambda)=K^t_{\bar{a}}(-\lambda-i3/2)$, where $K_a(\lambda)$ and $K_{\bar{a}}(\lambda)$  are a solution to the following boundary Yang-Baxter equations 
\bea
\fl  R_{ab}(\lambda-\mu)K_a(\lambda)R_{ba}(\lambda+\mu)K_b(\mu)=K_b(\mu)R_{ab}(\lambda+\mu)K_a(\lambda)R_{ba}(\lambda-\mu)\,,\label{eq:r1}\\
\fl \bar{R}_{ab}(\lambda-\mu)K_{\bar{a}}(\lambda)\bar{R}_{ba}(\lambda+\mu)K_b(\mu)=K_b(\mu)R_{ab}(\lambda+\mu)K_{\bar a}(\lambda){\bar R}_{ba}(\lambda-\mu)\,.\label{eq:r2}
\eea
A complete classification of the solutions to these equations can be found in \cite{AACD04}. In this work, we will not use this type of integrable boundary conditions, so that they will not be discussed further here; the interested reader is referred to \cite{AACD04} for more details.

\paragraph{Soliton non-preserving boundary conditions.} In the framework of quantum spin chains, soliton-non-preserving boundary conditions were first introduced in \cite{Doik00} for the $SU(3)$-invariant model, and later systematically studied for the $SU(\mathcal{N})$-invariant case in \cite{AACD04}. The soliton-non-preserving boundary transfer matrices are defined as
\bea
\tau_{\rm SNP}(\lambda)&=&{\rm tr}_{a}\left\{K_a^{+}(\lambda)T_a(\lambda)K_a^{-}(\lambda)\hat{T}_{\bar{a}}(\lambda)\right\}\,,\label{eq:def_tau}\\
\bar{\tau}_{\rm SNP}(\lambda)&=&{\rm tr}_a\left\{K_{\bar{a}}^{+}(\lambda)T_{\bar{a}}(\lambda)K_{\bar{a}}^{-}(\lambda)\hat{T}_a(\lambda)\right\}\,,\label{eq:def_bar_tau}
\eea
where now
\bea
 T_a(\lambda)=R_{aN}(\lambda-\xi_{N})\bar{R}_{a(N-1)}(\lambda-\xi_{N-1})\ldots R_{a2}(\lambda-\xi_2) \bar{R}_{a1}(\lambda-\xi_1)\,,\label{eq:def_t_a}\\
 \hat{T}_{\bar{a}}(\lambda)=R_{1a}(\lambda+\xi_1)\bar{R}_{2a}(\lambda+\xi_2)\ldots R_{(N-1)a}(\lambda+\xi_{N-1}) \bar{R}_{Na}(\lambda+\xi_{N})\,,\\
 T_{\bar{a}}(\lambda)=\bar{R}_{aN}(\lambda-\xi_{N})R_{a(N-1)}(\lambda-\xi_{N-1})\ldots \bar{R}_{a2}(\lambda-\xi_2) R_{a1}(\lambda-\xi_1)\,,\label{eq:def_t_bar_a}\\
 \hat{T}_{a}(\lambda)=\bar{R}_{1a}(\lambda+\xi_1)R_{2a}(\lambda+\xi_2)\ldots \bar{R}_{(N-1)a}(\lambda+\xi_{N-1}) R_{Na}(\lambda+\xi_{N})\,.
\eea
Once again, we introduced the inhomogeneities $\xi_j$, while $K_a^\pm(\lambda)$, $K_{\bar{a}}^\pm(\lambda)$ are $3\times 3$ matrices. In this case, they have to be chosen as $K_{a}^-(\lambda)=K_{a}(\lambda)$, $K_{\bar{a}}^-(\lambda)=K_{\bar{a}}(\lambda)$,  $K_{a}^+(\lambda)=K^t_{a}(-\lambda-i3/2)$ and $K_{\bar{a}}^+(\lambda)=K^t_{\bar{a}}(-\lambda-i3/2)$, where $K_a(\lambda)$ and $K_{\bar{a}}(\lambda)$ are a solution to the equations
\bea
 \fl R_{ab}(\lambda-\mu)K_a(\lambda)\bar{R}_{ba}(\lambda+\mu)K_b(\mu)=K_b(\mu)\bar{R}_{ab}(\lambda+\mu)K_a(\lambda)R_{ba}(\lambda-\mu)\,,\label{eq:reflection_1}\\
\fl \bar{R}_{ab}(\lambda-\mu)K_{\bar{a}}(\lambda)R_{ba}(\lambda+\mu)K_b(\mu)=K_b(\mu)R_{ab}(\lambda+\mu)K_{\bar a}(\lambda){\bar R}_{ba}(\lambda-\mu)\,.\label{eq:reflection_2}
\eea
Note that Eqs.~\eqref{eq:reflection_1}, \eqref{eq:reflection_2} differ from Eqs.~\eqref{eq:r1},\eqref{eq:r2} of the soliton-preserving case. To distinguish the two, the former are usually called twisted boundary Yang-Baxter equations. As a final remark, we note that one could also define the transfer matrices \eqref{eq:def_tau}, \eqref{eq:def_bar_tau} for spin-$1/2$ Heisenberg spin chains. However, one can see that in this case the latter would be equivalent to the soliton-preserving boundary matrices introduced above, due to the presence of the crossing relation \eqref{eq:crossin_relation} \cite{AACD04}.

\subsection{Identification between initial states and boundary parameters}\label{sec:identification}

Having introduced the integrable boundary transfer matrices that exist for $SU(3)$-invariant spin chains, our goal is now to relate either one of them to the operator $\mathcal{T}$ defined in \eqref{eq:mathcalT}.  As a first observation, we see that the latter is defined in terms of two ``quantum monodromy matrices'' $T^{\rm QTM}_j(\lambda)$, that contains alternating products of $R$-matrices $R_{i,j}(\lambda)$ and $\bar{R}_{i,j}(\lambda)$, cf. Eq.~\eqref{eq:q_monodromy}. Comparing with both the soliton-preserving and soliton-non-preserving boundary transfer matrices, it is clear that the operator $\mathcal{T}$ shares a similar structure only with the latter. In fact, the two operators can be proportional, provided that the initial state is chosen to match some appropriate $K$-matrix.

As a first step to identify $\mathcal{T}$ with an integrable boundary transfer matrix, we perform a shift in the inhomogeneities and the spectral parameter of the soliton-non-preserving operator $\tau_{\rm SNP}(\lambda)$; namely we use the redefinition
\be 
\lambda \to \lambda -3i/4\,, \qquad \xi_i \to  \xi_i^s=\xi_i +3i/4\,, 
\label{eq:shift_parameters}
\ee 
and a corresponding shift in the arguments of the $K_\pm$ matrices. Accordingly, we define the shifted transfer matrix
\be
\tau^s_{\rm SNP}(\lambda)={\rm tr}\left\{K^{+}(\lambda-3i/4)T^s_a(\lambda)K^{-}(\lambda-3i/4)\hat{T}^s_{\bar{a}}(\lambda)\right\}\,,
\label{eq:open_transfer_matrix}
\ee
where 
\bea
\fl T^s_a(\lambda)=R_{aN}(\lambda-\xi^s_N)\bar{R}_{a(N-1)}(\lambda-\xi^s_{N-1})\ldots R_{a2}(\lambda-\xi^s_2) \bar{R}_{a1}(\lambda-\xi^s_1)\,,\label{eq:shifted_monodromy}\\
\fl \hat{T}^s_{\bar{a}}(\lambda)=R_{1a}(\lambda+\xi^s_1-3i/2)\bar{R}_{2a}(\lambda+\xi^s_2-3i/2)\ldots 
\nonumber\\
R_{(N-1)a}(\lambda+\xi^s_{N-1}-3i/2) \bar{R}_{Na}(\lambda+\xi^s_N-3i/2)\,,
\eea
Note that with the same shift \eqref{eq:shift_parameters}, the transfer \eqref{eq:def_bar_tau} is rewritten as
\bea
\bar{\tau}^s_{\rm SNP}(\lambda)&=&{\rm tr}_a\left\{K_{\bar{a}}^{+}(\lambda-3i/4)T^s_{\bar{a}}(\lambda)K_{\bar{a}}^{-}(\lambda-3i/4)\hat{T}^s_a(\lambda)\right\}\,,\label{eq:bar_open_transfer_matrix}
\eea
where now
\bea
\fl T^s_{\bar{a}}(\lambda)=\bar{R}_{aN}(\lambda-\xi^s_{N})R_{a(N-1)}(\lambda-\xi^s_{N-1})\ldots \bar{R}_{a2}(\lambda-\xi^s_2) R_{a1}(\lambda-\xi^s_1)\,,\label{eq:shifted_monodromy_bar}\\
\fl \hat{T}^s_{a}(\lambda)=\bar{R}_{1a}(\lambda+\xi^s_1-3i/2)R_{2a}(\lambda+\xi^s_2-3i/2)\ldots \nonumber\\
\bar{R}_{(N-1)a}(\lambda+\xi^s_{N-1}-3i/2) R_{Na}(\lambda+\xi^s_{N}-3i/2)\,.
\eea
At this point it is straightforward to repeat the calculations reported in the case of $XXZ$ Heisenberg spin chains in \cite{PiPV17}. In particular, setting the inhomogeneities as
\bea 
\xi_{2i}^s &=&  - \mu =  \frac{i\beta}{N}\,, \label{eq:inhomogeneities_1} \\
\xi_{2i+1}^s &=&  - \bar{\mu} =  -\frac{i\beta}{N} + 3i/2\,,
\label{eq:inhomogeneities_2}
\eea 
 and choosing the two-site building block of the initial state as 
\bea
|\psi_0 \rangle_{1,2} &=& \sum_{i,j=1,3} (K^-(-3i/4)V )_{i,j} |i\rangle_1 |j\rangle_2\,,  
\label{eq:identification_1}\
\eea
where $V$ is defined in \eqref{eq:v_matrix}, one can easily verify
\be 
\mathcal{T} =\frac{1}{\langle \psi_0|\psi_0\rangle(1-\beta/2N)^{2N}}\tau_{\rm SNP}^s(0)\,.
\label{eq:key_relation}
\ee 

Eq.~\eqref{eq:key_relation} is an important relation: it tells us that the two-row transfer matrix $\mathcal{T}$ driving the evolution in the rotated picture (cf. Fig.~\ref{fig:2d_classical}) can be indeed equivalent (namely, proportional) to an appropriate soliton-non-preserving integrable boundary transfer matrix $\tau_{\rm SNP}(0)$ (in the following, we will omit the label SNP, and it will be understood that we will always work with this kind of boundary transfer matrix). Eq.~\eqref{eq:identification_1} instead defines the initial states for which such identification is possible. In the following section we will write them down explicitly, and show that they are integrable in the sense of Ref.~\cite{PiPV17_II}. These are the states whose quench dynamics is analyzed in this work. 

\subsection{The final result} 

We are now in a position to write down explicitly the integrable initial product states that correspond to integrable boundary conditions in the rotated channel. In order to do this, we start from Eq.~\eqref{eq:identification_1}, and use the general form of the $K$-matrix $K^-(\lambda)$ which satisfies the twisted boundary Yang-Baxter equations \eqref{eq:reflection_1}, \eqref{eq:reflection_2}.
In Ref.\cite{AACD04}, it was shown that, in the $SU(3)$ case, the only invertible solutions to the latter equations are scalar matrices $K^-(\lambda)=K^-$ such that $(K^-)^{t} = K^-$, namely such that $\tilde{K}^- = K^- V$ is symmetric. The most general matrix of this form is written as
\be
K_a(\lambda)=K_sV\,,
\label{eq:K_symmetric}
\ee
with $K_{s}$ a symmetric numerical $3\times 3$ matrix
\be
K_{s}=\left(
\begin{array}{ccc}
	\kappa_{11} & \kappa_{12} & \kappa_{13}\\
	\kappa_{12} & \kappa_{22} & \kappa_{23}\\
	\kappa_{13} & \kappa_{23} & \kappa_{33}
\end{array}  
\right),
\ee
with ${\rm det}K_s\neq 0$.  Plugging this into \eqref{eq:identification_1} we obtain 
\bea 
| \psi_0 \rangle &=& \kappa_{11} |1,1\rangle +\kappa_{22} |2,2\rangle +  \kappa_{33} |3,3\rangle 
+ \kappa_{12}( |1,2\rangle +  |2,1\rangle ) \nonumber\\
&+& \kappa_{13}( |1,3\rangle +  |3,1\rangle ) 
+ \kappa_{23}( |2,3\rangle +  |3,2\rangle )\,. 
\label{eq:integrable_block}
\eea  
This is our final result for the boundary states of the model, which
are those for which the quench dynamics will be analyzed in this
paper. Namely, we will consider the product states \eqref{eq:product_states} with the two-site block of the form \eqref{eq:integrable_block}. Note that this class does not contain N\'eel-like states, such as in the XXZ Heisenberg chain. Furthermore, there is a distinguished state in this class, where the
  diagonal components in the standard basis are equal; this will be
  called the ``delta-state'':
  \be
  | \psi_\delta \rangle = \frac{1}{\sqrt{3}}\left(|1,1\rangle + |2,2\rangle + |3,3\rangle \right)\,.
  \label{eq:deltastate}
  \ee
This state is invariant with respect to $SO(3)$ rotations, and this symmetry will manifest itself in the exact solution of the quench. Note that the coefficients $\kappa_{ij}$ can be chosen arbitrarily, so that one has a large freedom in the choice of the initial state.

In Ref.~\cite{PiPV17_II} integrable states were defined as those matrix product states with finite bond dimension which are annihilated by the charges that are odd under space reflection. So far, we have only identified product states that can be related to integrable boundaries in the rotated channel. However, in analogy to \cite{PiPV17_II}, it is possible to prove that such states are indeed annihilated by the odd charges. This is proved in \ref{sec:proof_of_integrability}. More precisely, there we show that product states \eqref{eq:product_states} with the two-site block derived above satisfy
\be 
t(\lambda) | \Psi_0 \rangle = \Pi t(\lambda) \Pi | \Psi_0 \rangle \,.
\label{eq:starrel}
\ee
Here $t(\lambda)$ is the periodic transfer matrix \eqref{eq:periodic_transfer_matrix}, while $\Pi$ is the parity operator on the spin-chain, which is defined by its action on the basis vectors
\be
\Pi |i_1,\ldots,i_L\rangle =|i_L,\ldots,i_1 \rangle\,.
\ee
The proof generalizes the one outlined in \cite{PiPV17_II}, but does not make use of the crossing relation of the $R$-matrix, which is not present in the $SU(3)$-invariant spin chain. As a consequence of \eqref{eq:starrel}, repeating the reasoning reported in \cite{PiPV17_II}, one immediately gets that boundary states $|\Psi_0\rangle$ are annihilated by all odd charges of the model, and hence are integrable according to the definition of \cite{PiPV17_II}. 

It is interesting to mention that one could also find product states which satisfy \eqref{eq:starrel} but do not correspond to invertible $K$-matrices. These states are in fact related to the $SU(2)$-integrable states and are thus not discussed further here. The interested reader is referred to \ref{sec:non-invertible_K} for more discussions on this point.

Finally, following \cite{PiPV17_II}, we note that, starting from the integrable states derived above, an additional infinite number of integrable matrix product states can be straightforwardly built by subsequent application of (fused) transfer matrices to $|\Psi_0\rangle$. Even though these states might display interesting features, they will not be considered in this work.

\subsection{Restriction to diagonal reflection matrices}

In the following, we will focus on the class of integrable product states \eqref{eq:product_states} where the two-site block  \eqref{eq:integrable_block} is chosen to contain only diagonal terms, namely
\bea 
| \psi_0 \rangle &=& \kappa_{11} |1,1\rangle +\kappa_{22} |2,2\rangle +  \kappa_{33} |3,3\rangle\,,
\label{eq:integrable_block_diagonal}
\eea
further restricting ourselves to the case
\be
|\kappa_{11}|\geq|\kappa_{22}|\geq |\kappa_{33}|\,.
\label{eq:maximal_magnetization}
\ee
From the analytical point of view, the states
\eqref{eq:integrable_block_diagonal} are easier to deal with, if
compared to the general integrable states
\eqref{eq:integrable_block}, but there are other reasons to focus on \eqref{eq:integrable_block_diagonal}.

First, in \ref{sec:restriction_K_matrices} it is shown that all the
integrable states presented in the previous section can be related to
the case of diagonal $K$-matrices via a global rotation $W\in
SU(3)$; this is achieved by a Schmidt decomposition. Hence, if one is interested in the dynamics of $SU(3)$-invariant operators, it is not a restriction to consider only the case \eqref{eq:integrable_block_diagonal}. 

The second reason to restrict ourselves to
\eqref{eq:integrable_block_diagonal} can be better understood
thinking about the case of quantum quenches in the $SU(2)$-invariant
Heisenberg spin chain. Indeed, it is by now well established that in
this model the quasilocal charges invariant under spin-inversion \cite{IMPZ16} fully determine the rapidity
distribution functions of the post-quench steady states
\cite{IQNB16}. However, these charges are invariant under the global
action of $SU(2)$, so that states that are related via a
$SU(2)$-rotation correspond to the same rapidity distribution
functions. In this case, the thermodynamic description only differs by
the density of rapidities at infinity \cite{DWBC14_II}. It is natural to conjecture that a similar scenario takes place also
for quenches in the $SU(3)$-invariant spin chains, and that states that
are related via a global $SU(3)$ rotation correspond to the same
rapidity distribution functions. Under this hypothesis, such distribution functions for a state \eqref{eq:integrable_block} can always be
obtained by computing the one for the diagonal state related by a global $SU(3)$ rotation. Finally, the condition
\eqref{eq:maximal_magnetization} amounts to choosing the minimum
density of magnons between the class of diagonal states. In fact, we
will see that the post-quench steady state corresponding to these
initial states displays vanishing densities of rapidities at
infinity.

\section{The post-quench steady states}\label{sec:distribution_functions}

In this section we tackle the problem of characterizing the stationary state reached at large times after a quench from integrable initial states. This can be done by providing the corresponding rapidity distribution functions of the quasi-particles $\rho_{n}^{(r)}(\lambda)$, which are widely believed to encode all the macroscopic information of a given state. As we discussed in Sec.~\ref{sec:intro1}, one of the merits of the Quench Action approach has been to clarify how a description in terms of rapidity distribution functions emerges for the post-quench steady state. However, the actual derivation of the latter using this approach necessarily requires the knowledge of the overlaps between the initial state and the eigenstates of the model, which are difficult to compute.

In this work, we follow a different strategy to derive the post-quench rapidity distribution functions, which was first employed in \cite{PiPV17} for XXZ Heisenberg chains. This approach is based on a comparison between the QTM formalism and the Quench Action approach, and the (well-motivated) assumption that the functions $\eta_n^{(r)}(\lambda)$ defined in \eqref{eq:eta_functions} coincide with certain $Y$-functions that arise naturally in the QTM construction, and that can be analytically computed.

In the next subsection, we begin by recalling the main ideas of the Quench Action method, which are going to be helpful to better explain the QTM derivation of the post-quench rapidity distribution functions. The latter will in fact follow the one already reported in \cite{PiPV17} for XXZ Heisenberg chains. Note that our derivation is not based on the knowledge of the overlaps between the initial states and the eigenstates of the model, but instead employs certain ``fusion relations'' of the soliton-non-preserving boundary transfer matrices. In the case of $SU(2)$-invariant spin chains, such relations were derived in \cite{Zhou95,Zhou96}. A generalization to the $SU(3)$-case is non-trivial, and to our knowledge not yet present in the literature. Since their derivation goes beyond the scope of our work, the latter is reported in \cite{InPrep}, while here we simply use the final results of the analysis reported therein.  

\subsection{The Quench Action method}
\label{sec:QAM}
In order to introduce the main ideas of the Quench Action method, let us consider the computation of the partition function \eqref{eq:euclidean_partition_function}
\be
\langle \Psi_0|e^{-\beta H_L}| \Psi_0\rangle=\sum_{n}|\langle \Psi_0|n\rangle|^2e^{-\beta E_n}\,.
\label{eq:finite_sum}
\ee
Here the sum is over all the eigenstates of the Hamiltonian, whose energy is indicated with $E_n$. For integrable models eigenstates are labeled by sets of rapidities; in particular, in our case the sum is over all possible sets $\{\lambda_j\}$, $\{\mu_j\}$ that are a solution to the Bethe equations \eqref{eq:bethe_equations1}, \eqref{eq:bethe_equations2}. In analogy with the computation of thermal partition functions within the thermodynamic Bethe ansatz approach \cite{takahashi-99}, in the limit of large system sizes $L$, we can replace the sum with a functional integral over all possible rapidity distribution functions $\{\rho_n^{(r)}(\lambda)\}$. The first step in the application of the Quench Action approach then relies in the possibility of defining a functional $S_{Q}$ such that in the large-$L$ limit
\be
|\langle \Psi_0|\{\lambda_j\},\{\mu_k\}\rangle|^2\simeq \exp\left(-LS_Q\left[\{\rho_n^{(r)}(\lambda)\}\right]\right)
\,,
\label{eq:overlap_SQ}
\ee
where we denoted with $|\{\lambda_j\},\{\mu_k\}\rangle$ the eigenstate corresponding to the sets $\{\lambda_j\},\{\mu_k\}$ (the latter arranging themselves according to the distributions $\{\rho_n^{(r)}(\lambda)\}$ in the thermodynamic limit). Note that the existence of such a functional is a priori non-trivial, as Eq.~\eqref{eq:overlap_SQ} implies that the overlap between $|\Psi_0\rangle$ and the eigenstates of the Hamiltonian only depends on the set $\{\rho_n^{(r)}(\lambda)\}$ and not on the details of the specific eigenstate. Assuming the existence of such a functional, in the large-$L$ limit one can rewrite Eq.~\eqref{eq:finite_sum} as
\be
\langle \Psi_0|e^{-\beta H}|\Psi_0\rangle=\int \mathcal{D}\,\boldsymbol{\rho}\exp\{-\beta\,e[\boldsymbol{\rho}]L+S_{\rm QA}[\boldsymbol{\rho}]L\}\,.
\label{eq:functional_integral}
\ee
Here we introduced the Quench Action functional
\be 
S_{\rm QA}[{\boldsymbol{\rho}}]=2S_Q[{\boldsymbol{\rho}}]-\frac{1}{2}S_{\rm YY}[{\boldsymbol{\rho}}]\,,
\label{eq:SQA}
\ee
where ${\boldsymbol \rho}$ denotes the sets $\{\rho^{(r)}_n(\lambda)\}_{n=1}^{\infty}$ and where the Yang-Yang entropy \cite{YaYa69,johannesson-86}
\bea
S_{\rm YY}\left[{\boldsymbol{\rho}}\right]&=&\sum_{r=1}^{2}\sum_{n=1}^{+\infty}\int_{-\infty}^{+\infty}{\rm d}x \left\{\left(\rho_n^{(r)}(x)+\rho_{h,n}^{(r)}(x)\right)\ln\left(\rho_n^{(r)}(x)+\rho_{h,n}^{(r)}(x)\right)\right.\nonumber\\
&-&\left.\rho_{n}^{(r)}(x)\ln\rho_{n}^{(r)}(x)-\rho_{h,n}^{(r)}(x)\ln\rho_{h,n}^{(r)}(x)
\right\}\,,
\label{eq:yangyang}
\eea
takes into account that to each set of rapidity distribution functions $\{\rho_n^{(r)}(\lambda)\}$ are associated an exponentially large number of eigenstates. Note that, differently from the thermal case \cite{takahashi-99}, we have a factor $1/2$ in front of the Yang-Yang entropy in \eqref{eq:SQA}, which comes from the fact that integrable initial states $|\Psi_0\rangle$ have by definition non-vanishing overlaps only with microscopically parity-symmetric Bethe states \cite{DWBC14,PiPV17}. Finally, $e[\boldsymbol{\rho}]$ is the energy density defined in Eq.~\eqref{eq:energy_density}.
The functional integral \eqref{eq:functional_integral} can now be computed by saddle-point evaluation, leading to the condition
\be
\frac{\partial \left(-\beta\,e[\boldsymbol{\rho}]+S_{\rm QA}[\boldsymbol{\rho}]\right)}{\partial \rho^{(r)}_n(\lambda)} =0 \qquad\qquad r=1,2\,,\qquad\qquad n\geq 1\,.
\label{eq:general_saddle_point}
\ee

Assuming the validity of \eqref{eq:overlap_SQ}, and by means of a slightly more sophisticated argument, one can in fact derive a much more interesting relation for the large-time limit of a local observable $\mathcal{O}$ after a quench \cite{CaEs13}, namely
\be
\lim_{t\to\infty}\lim_{L\to\infty}\langle \Psi_0|\mathcal{O}(t)|\Psi_0\rangle=\langle {\boldsymbol \rho}_{s} |\mathcal{O}|{\boldsymbol \rho_{s}} \rangle\,.
\label{eq:observables}
\ee
Here $|\boldsymbol{\rho_{s}}\rangle$ is an eigenstate corresponding to the set of rapidity distribution functions $\boldsymbol{\rho_{s}}$ , which in turn are obtained as the solution to the saddle-point equations 
\be
\frac{\partial \left(S_{\rm QA}[\boldsymbol{\rho}_s]\right)}{\partial \rho^{(r)}_n(\lambda)} =0 \qquad\qquad r=1,2\,,\qquad\qquad n\geq 1\,.
\label{eq:saddle_point}
\ee
Eq.~\eqref{eq:observables} is a key relation of the QAM: it tells us that for the purpose of computing local properties, the system at large times can be successfully described by a representative eigenstate characterized by the set of distributions $\boldsymbol{ \rho_{s}}$. Comparing Eq.~\eqref{eq:saddle_point} with Eq.~\eqref{eq:general_saddle_point}, we see that the post-quench rapidity distribution functions $\boldsymbol{\rho}_s$ can be obtained by computing the partition function \eqref{eq:functional_integral}, and eventually taking the limit $\beta\to 0$ of the solution to the saddle-point equations \eqref{eq:general_saddle_point}.

One can go on, and elaborate further on the saddle-point condition \eqref{eq:general_saddle_point}. Indeed, under mild assumptions on the structure of the overlaps, Eq.~\eqref{eq:general_saddle_point} can be cast into a set of non-linear integral equations for the functions $\eta_j^{(r)}(\lambda)$, namely \cite{MBPC17}
\bea
\ln \eta_{n}^{(1)}(\lambda)&=&2\beta\varepsilon_n(\lambda)-g^{(1)}_n(\lambda)+\sum_{m=1}^{+\infty}\left[a_{n,m}\ast \ln\left(1+\left[\eta_m^{(1)}\right]^{-1}\right)\right](\lambda)\nonumber\\
&-&\sum_{m=1}^{+\infty}\left[b_{n,m}\ast \ln\left(1+\left[\eta_m^{(2)}\right]^{-1}\right)\right](\lambda)\,,\label{eq:general_eta_1}\\
\ln \eta_{n}^{(2)}(\lambda)&=&-g^{(2)}_n(\lambda)+\sum_{m=1}^{+\infty}\left[a_{n,m}\ast \ln\left(1+\left[\eta_m^{(2)}\right]^{-1}\right)\right](\lambda)\nonumber\\
&-&\sum_{m=1}^{+\infty}\left[b_{n,m}\ast \ln\left(1+\left[\eta_m^{(1)}\right]^{-1}\right)\right](\lambda)\,,\label{eq:general_eta_2}
\eea
where the functions $a_{n,m}$, $b_{n,m}$ and $\varepsilon_n(\lambda)$ are defined in \eqref{eq:a_mn}, \eqref{eq:b_mn} and \eqref{eq:epsilon_energy} respectively, while $g_{n}^{(r)}(\lambda)$ are determined completely by the overlaps. Note that the above system has the traditional form of the thermodynamic Bethe ansatz equations arising at thermal equilibrium for $SU(3)$-invariant spin chains \cite{mntt-93}. In particular, it can be rewritten in a partially decoupled form as (see e.g. \cite{MBPC17})
\bea
\fl \ln \eta_m^{(1)}(\lambda) =-4\pi \beta s(\lambda)\delta_{m,1} + \left[s \ast \ln\left(  \frac{(1+\eta_{m+1}^{(1)})(1+\eta_{m-1}^{(1)})}{1+(\eta_m^{(2)})^{-1}} \right)\right](\lambda) + h^{(1)}_m(\lambda)\,, \label{eq:TBA_1}\\
\fl \ln \eta_m^{(2)}(\lambda) = \left[s \ast \ln\left(  \frac{(1+\eta_{m+1}^{(2)})(1+\eta_{m-1}^{(2)})}{1+(\eta_m^{(1)})^{-1}} \right)\right](\lambda) + h^{(2)}_m(\lambda)\,, \label{eq:TBA_2}
\eea
with the convention $\eta_{0}^{(r)}(\lambda)\equiv 0$, and where $h^{(r)}_m(\lambda)$ are appropriate driving terms which can be derived from the knowledge of $g_{n}^{(r)}(\lambda)$. 

Once the functions $\eta_j^{(r)}(\lambda)$ are obtained as the solution to the above equations, one can solve the thermodynamic form of the Bethe equations \eqref{eq:TBAexplicit1}, \eqref{eq:TBAexplicit2} and finally obtain the rapidity distribution functions $\rho^{(r)}_n(\lambda)$ of the post-quench steady state. Since the driving terms of the above integral equations depend on the overlaps between the initial state and the eigenstates of the model, the latter are apparently needed in order to determine the post-quench rapidity distribution functions. In the following, however, we provide a derivation where a knowledge of the overlaps is not used. Indeed, the partition function \eqref{eq:euclidean_partition_function} can be derived independently using the QTM approach, and is expressed in terms of certain ``$Y$-functions'' which solve a system of integral equations of the same form of \eqref{eq:TBA_1} and \eqref{eq:TBA_2} (see Ref.~\cite{InPrep} where this is done explicitly)). Then, we conjecture that one can identify such $Y$-functions with the distributions $\eta^{(r)}_j(\lambda)$, obtaining a derivation of the latter which is independent of the QAM. Even though such a conjecture is non-trivial, it is very natural and in the XXZ Heisenberg chain was verified analytically in \cite{PiPV17} for all integrable quenches. In this work, we will assume that this conjecture holds true also for the $SU(3)$-invariant spin chain. We will present later a number of analytic tests of our final results, which will provide strong evidence supporting this assumption.

\subsection{Derivation of the post-quench rapidity distribution functions}

The QTM derivation of the post-quench rapidity distribution functions begins with the analysis of the partition function \eqref{eq:euclidean_partition_function}. It follows from Eqs.~\eqref{eq:ST_dec}, \eqref{eq:rhoNLQTM} that in the large-$L$ limit, the latter can be obtained by computing the leading eigenvalue of the two-row transfer matrix $\mathcal{T}$ (cf. Ref.~\cite{PiPV17}). Hence, using the identification \eqref{eq:key_relation} one is left with the problem of computing the leading eigenvalue of an integrable boundary transfer matrix with soliton-non-preserving boundary conditions. In the case of $XXZ$ Heisenberg spin chains, a particularly convenient method was employed in \cite{PiPV17} based on the existence of some fusion relations of boundary transfer matrices. The same approach is followed here.

First, the transfer matrices $\tau^s(\lambda)$, $\bar{\tau}^s(\lambda)$ can be embedded into an infinite family of ``fused'' transfer matrices $\tau^{(r)}_n(\lambda)$, with $r=1,2$ and $n=1,\ldots =\infty$, such that  $\tau^{(1)}_1(\lambda)=\tau^s(\lambda)$ and $\tau^{(2)}_1(\lambda)=\bar{\tau}^s(\lambda)$ (the name of such matrices comes from the fact that they can be built through an appropriate fusion procedure \cite{KuRS81} starting from $\tau^s(\lambda)$ and $\bar{\tau}^s(\lambda)$). The boundary transfer matrices $\tau^{(r)}_n(\lambda)$ act on the same Hilbert space of $\tau^s(\lambda)$ and are defined in analogy with \eqref{eq:def_tau}. However for $\tau^{(1)}_n(\lambda)$ and $\tau^{(2)}_n(\lambda)$ the trace is taken over a $d$-dimensional space, with $d=(n+1)(n+2)/2$, which corresponds to the $n$-fold symmetric tensor of the $SU(3)$-fundamental and conjugate representations respectively. Importantly, the fused transfer matrices commute, namely
\be
\left[\tau^{(r)}_n(\lambda),\tau^{(s)}_m(\lambda)\right]=0\,.
\label{eq:commutation}
\ee
Note that, due to \eqref{eq:commutation}, one can choose a common basis of eigenvectors for $\tau^{(r)}_n(\lambda)$. 

In the case of $XXZ$ Heisenberg spin chains, a set of functional relations relating the eigenvalues of the fused transfer matrices was derived in \cite{Zhou95,Zhou96}.  A generalization to the $SU(3)$-case and soliton-non-preserving boundary conditions is non-trivial, and to our knowledge not yet present in the literature. We report the derivation of such fusion relations in our second paper \cite{InPrep}, where further discussions on the fused transfer matrices are reported, while in the following we will only need the final result of the analysis of \cite{InPrep}, which is reported below.

It is derived in \cite{InPrep} that the eigenvalues of the fused transfer matrices are related by the functional relations \cite{InPrep}
\bea
\tau^{(1)}_m\left(u+\frac{i}{2}\right)\tau^{(1)}_m\left(u-\frac{i}{2}\right)&=&\tau^{(1)}_{m+1}(u)\tau^{(1)}_{m-1}(u)+	\Phi^{(1)}_{m}(u)\tau^{(2)}_{m}(u)\,, \label{eq:boundary_t_system1} \\ 
\tau^{(2)}_m\left(u+\frac{i}{2}\right)\tau^{(2)}_m\left(u-\frac{i}{2}\right)&=&\tau^{(2)}_{m+1}(u)\tau^{(2)}_{m-1}(u)+\tilde{\Phi}^{(2)}_{m}(u)\tau^{(1)}_{m}(u)\,,
\label{eq:boundary_t_system2}
\eea
with the identifications
\bea
\tau_0^{(1)}(u)&\equiv& \tau_0^{(2)}(u)\equiv 1\,,\\
\tau_1^{(1)}(u)&=&\tau^s(u)\,,\\
\tau_1^{(2)}(u)&=&\bar{\tau}^s(u)\,.
\eea
Here we introduced
\bea
\Phi_n^{(1)}(\lambda)=\prod_{j=1}^{n}f^{(1)}\left[\lambda-(n-2j+1)\frac{i}{2}\right]\,,\label{eq:final_result_1}\\
\Phi_n^{(2)}(\lambda)=\prod_{j=1}^{n}f^{(2)}\left[\lambda-(n-2j+1)\frac{i}{2}\right]\,,\label{eq:final_result_2}
\eea
where
\bea
f^{(1)}(\lambda)&=&-\frac{4\lambda^2}{\zeta(2\lambda)}\left[\prod_{j=1}^{N/2}\left(i/2-\lambda+\xi_{2j}\right)\left(2i+\lambda-\xi_{2j-1}\right)\right.\nonumber\\
&\times&\left.\left(2i-\lambda-\xi_{2j-1}\right)\left(i/2+\lambda+\xi_{2j}\right)\right]\left[{\rm det}K_s\right]^2\,,\label{eq:final_result_3}\\
f^{(2)}(\lambda)&=&-\frac{4\lambda^2}{\zeta(2\lambda)}\left[\prod_{j=1}^{N/2}\left(i/2-\lambda+\xi_{2j-1}\right)\left(2i+\lambda-\xi_{2j}\right)\right.\nonumber\\
&\times&\left.\left(2i-\lambda-\xi_{2j}\right)\left(i/2+\lambda+\xi_{2j-1}\right)\right]\left[{\rm det}K_s\right]^{-2}\,.\label{eq:final_result_4}
\eea
The functional equations \eqref{eq:boundary_t_system1} and \eqref{eq:boundary_t_system2} constitute the so-called $T$-system of boundary transfer matrices. Its usefulness lies in the possibility of casting it into a system of integral equations for $\tau^{(r)}_n(\lambda)$, which can be eventually solved numerically. In this way, one obtains the function $\tau^s(\lambda)$ and hence, from \eqref{eq:key_relation} the value of $\mathcal{T}$. In order to proceed with this program, one needs to derive from \eqref{eq:boundary_t_system1} and \eqref{eq:boundary_t_system2} an equivalent set of functional equations which constitute  the  ``$Y$-system'' \cite{KuNS11}. In particular, setting
\bea
y_{j}^{(1)}(u)&=&\frac{\tau^{(1)}_{j+1}\left(u\right)\tau^{(1)}_{j-1}\left(u\right)}{\Phi^{(1)}_j(u)\tau^{(2)}_{j}\left(u\right)}\,,\label{eq:y1}\\
y_{j}^{(2)}(u)&=&\frac{\tau^{(2)}_{j+1}\left(u\right)\tau^{(2)}_{j-1}\left(u\right)}{\Phi^{(2)}_j(u)\tau^{(1)}_{j}\left(u\right)}\,, \label{eq:y2}
\eea 
one readily derives
\bea
y^{(1)}_j\left(u+\frac{i}{2}\right)y^{(1)}_j\left(u-\frac{i}{2}\right)&=&\frac{\left[1+y^{(1)}_{j-1}(u)\right]\left[1+y^{(1)}_{j+1}(u)\right]}{1+\left[y^{(2)}_{j}(u)\right]^{-1}}\,,\label{eq:y_system1}\\
y^{(2)}_j\left(u+\frac{i}{2}\right)y^{(2)}_j\left(u-\frac{i}{2}\right)&=&\frac{\left[1+y^{(2)}_{j-1}(u)\right]\left[1+y^{(2)}_{j+1}(u)\right]}{1+\left[y^{(1)}_{j}(u)\right]^{-1}}\,,\label{eq:y_system2}
\eea
with the convention $y^{(a)}_0(u)\equiv 0$. We note that the $Y$-system is a fundamental relation of integrable models \cite{KuNS11}, appearing repeatedly, for instance, in the context of thermal physics \cite{Suzu99,TaSK01,Tsub03} and, more recently, in quench problems \cite{DWBC14_II,PMWK14,PiCE16}.

We are now in a position to directly compare the QTM and Quench Action approaches. The comparison goes along the very same lines of the treatment in the XXZ Heisenberg spin chain carried out in \cite{PiPV17}. First, we notice that, using standard techniques in complex analysis, the $Y$-system can be rewritten in terms of non-linear integral equations as follows.  First, one takes the logarithmic derivative on both sides of Eqs.~\eqref{eq:y_system1}, \eqref{eq:y_system2} and Fourier transforms them. Then, the integrals appear in the l.h.s. along segments with non-zero imaginary parts: these can be moved to the real line, taking care of the poles of the logarithmic derivatives in the ``physical strip'' 
\be
\mathcal{S}=\left\{\lambda\left|-\frac{1}{2}\leq{\rm Im}(\lambda)\leq \frac{1}{2}\right.\right\}\,.
\label{eq:physical_strip}
\ee
At the end of this procedure one can transform back to real space, and obtain a set of non-linear integral equations of the exact same form of Eqs.~\eqref{eq:TBA_1}, \eqref{eq:TBA_2}. In this case, the driving terms $h^{(r)}_m(\lambda)$ are uniquely determined by the singularities of the logarithmic derivatives in the physical strip. The integral equations thus obtained can be solved for $y_j^{(r)}(\lambda)$. Inverting the functional relation \eqref{eq:y1} with analogous techniques, this result also gives the value of $\tau^s(\lambda)$, and hence of the partition function \eqref{eq:euclidean_partition_function}, through the identification \eqref{eq:key_relation}.

In summary, we get that there exist two ways of expressing the partition function \eqref{eq:euclidean_partition_function} in the thermodynamic limit, either in terms of $\eta^{(r)}_j(\lambda)$ or in terms of $y^{(r)}_j(\lambda)$. In order to compare the two sets of functions, one would need to compute explicitly the driving terms $h^{(r)}_m(\lambda)$. In the case of $XXZ$ Heisenberg chains the two systems of integral equations were shown to be identical for integrable quenches \cite{PiPV17} so that one gets the identification $\eta_j^{(r)}(\lambda)=y_j^{(r)}(\lambda)$. We conjecture that the same is true in the $SU(3)$-invariant case, namely that one can identify the functions $\eta^{(r)}_j(\lambda)$ as the solution to the $Y$-system in Eqs.~\eqref{eq:y_system1}, \eqref{eq:y_system2}.
 
Under this conjecture, one is left with the problem of finding a solution to the $Y$-system in the limit $\beta\to 0$. This can be done directly following the analogous calculation performed in Ref.~\cite{PiPV17} in the case of $XXZ$ Heisenberg chains . The details of this calculation for the $SU(3)$-Y-system are reported in our second paper \cite{InPrep}, together with the generic numerical solution in the case of finite $\beta$ . Putting all together, one arrives at the final result for the functions $\eta^{(r)}_j(\lambda)$. In particular, given an integrable initial state with a diagonal two-site building block \eqref{eq:integrable_block_diagonal} one has
\bea
\fl \eta_{1}^{(1)}(\lambda)=\frac{(|\kappa_{11}|^2+|\kappa_{11}|^2+|\kappa_{33}|^2)^2}{(1/|\kappa_{11}|^2+1/|\kappa_{22}|^2+1/|\kappa_{33}|^2)|\kappa_{11}\kappa_{22}\kappa_{33}|^2}\frac{\lambda^2+1/4}{\lambda^2}-1\,,\label{eq:analytical_eta1}\\
\fl \eta_{1}^{(2)}(\lambda)=\frac{((1/|\kappa_{11}|^2+1/|\kappa_{22}|^2+1/|\kappa_{33}|^2)^2|\kappa_{11}\kappa_{22}\kappa_{33}|^2}{|\kappa_{11}|^2+|\kappa_{11}|^2+|\kappa_{33}|^2}\frac{\lambda^2+1/4}{\lambda^2}-1\,,\label{eq:analytical_eta2}
\eea
while higher functions are obtained recursively from the formula
\bea
\eta_{n}^{(1)}(\lambda)&=&\frac{\eta^{(1)}_{n-1}(\lambda+i/2)\eta^{(1)}_{n-1}(\lambda-i/2)\left[1+\left(\eta_{n-1}^{(2)}\right)^{-1}\right]}{1+\eta_{n-2}^{(1)}(\lambda)}-1\,,\\
\eta_{n}^{(2)}(\lambda)&=&\frac{\eta_{n-1}^{(2)}(\lambda+i/2)\eta^{(2)}_{n-1}(\lambda-i/2)\left[1+\left(\eta_{n-1}^{(1)}\right)^{-1}\right]}{1+\eta_{n-2}^{(2)}(\lambda)}-1\,,
\eea
with the convention $\eta_{0}^{(1)}(\lambda)=\eta_{0}^{(2)}(\lambda)=0$. Note that these functions only depend on the absolute values of the entries $\kappa_{jj}$. Furthermore, they are expressed as simple rational functions of the rapidities $\lambda$, in analogy to the case of integrable states in the Heisenberg chain \cite{DWBC14_II,PiVC16}. Note that in the special case of the delta-state \eqref{eq:deltastate} we find that
\be
\eta_{1}^{(1)}(\lambda)=\eta_{1}^{(2)}(\lambda)=
\frac{3}{\lambda^2}\left(\lambda^2+\frac{1}{4}\right)-1\,.
\label{eq:analytical_etadelta}
\ee
It follows that a permutation symmetry $\eta_{j}^{(1)}(\lambda)=\eta_{j}^{(2)}(\lambda)$ will hold for all
higher functions $\eta_j^{(r)}(\lambda)$.

\subsection{The rapidity distribution functions}

In the previous subsection, we have derived the functions $\eta^{(r)}_j(\lambda)$ corresponding to the steady state reached at large times after a quench from an integrable state. Using this result, one can obtain, at least numerically, the rapidity distribution functions $\rho^{(r)}_j(\lambda)$: indeed, recalling that $\rho^{(r)}_{t,n}(\lambda)=\rho^{(r)}_{n}(\lambda)(1+\eta^{(r)}_j(\lambda))$, Eqs.~\eqref{eq:TBAexplicit1} and \eqref{eq:TBAexplicit2} can be solved iteratively for $\rho^{(r)}_{n}(\lambda)$, once $\eta^{(r)}_j(\lambda)$ are known. However, we find that it is possible to derive a fully analytic expression for $\rho^{(r)}_{n}(\lambda)$, based once again on a QTM construction explained in the following.

Consider the modified partition function
\be
\mathcal{Z}(\beta,s)=\langle \Psi_0|\bar{t}\left(s\right)t\left(s\right)e^{-\beta H_L}|\Psi_0\rangle\,,
\label{eq:modified_partition_function}
\ee
where $t(\lambda)$, $\bar{t}(\lambda)$ are the periodic transfer matrices defined in Eqs.~\eqref{eq:periodic_transfer_matrix} and \eqref{eq:periodic_bar_transfer_matrix}. Importantly, in the limit of small values of  $s$ we have \cite{FuKl99}
\be
\bar{t}\left(s\right)t\left(s\right)\simeq e^{-2is H_L}\,,
\label{eq:t_t_bar_h}
\ee
where $H_L$ is the Hamiltonian \eqref{eq:hamiltonian}. One could carry out a Quench-Action analysis along the lines of Sec.~\ref{sec:QAM}. Due to \eqref{eq:t_t_bar_h}, Eqs.~\eqref{eq:general_eta_1} and \eqref{eq:general_eta_2} would now be replaced by
\bea
\fl \ln \eta_{n,s}^{(1)}(\lambda)=\left[-4is+O(s^2)+2\beta\right]\varepsilon_n(\lambda)-g^{(1)}_n(\lambda)+\sum_{m=1}^{+\infty}\left[a_{n,m}\ast \ln\left(1+\left[\eta_{m,s}^{(1)}\right]^{-1}\right)\right](\lambda)\nonumber\\
-\sum_{m=1}^{+\infty}\left[b_{n,m}\ast \ln\left(1+\left[\eta_{m,s}^{(2)}\right]^{-1}\right)\right](\lambda)\,,\label{eq:eta_n_s_1}\\
\fl  \ln \eta_{n,s}^{(2)}(\lambda)=-g^{(2)}_n(\lambda)+\sum_{m=1}^{+\infty}\left[a_{n,m}\ast \ln\left(1+\left[\eta_{m,s}^{(2)}\right]^{-1}\right)\right](\lambda)\nonumber\\
-\sum_{m=1}^{+\infty}\left[b_{n,m}\ast \ln\left(1+\left[\eta_{m,s}^{(1)}\right]^{-1}\right)\right](\lambda)\,.\label{eq:eta_n_s_2}
\eea
For a fixed value of $\beta$, the functions $\eta_{n,s}^{(r)}(\lambda)$ now depend on the additional parameter $s$. In fact, it is easy to prove that
\bea
\rho_{t,j}^{(r)}(\lambda)=-\frac{i}{8\pi}\frac{\partial}{\partial s}\log\left[\eta_{j,s}^{(r)}(\lambda)\right]\Big|_{s=0}\,.
\label{eq:implicit_rho_j}
\eea
In order to do so, one simply notices that the system of equations for the r.h.s. of Eq.~\eqref{eq:implicit_rho_j}, which is obtained by differentiating both sides of Eqs.~\eqref{eq:eta_n_s_1}, \eqref{eq:eta_n_s_2}, coincides with the one for the  functions $\rho^{(r)}_{t,j}(\lambda)$ in Eqs.~\eqref{eq:TBAexplicit1} and \eqref{eq:TBAexplicit2}.

From Eq.~\eqref{eq:implicit_rho_j} we see that in order to obtain an analytic expression for $\rho_{t,j}^{(r)}(\lambda)$ it is enough to determine $\eta_{j,s}^{(r)}(\lambda)$; this can be done in complete analogy with the computation of $\eta^{(r)}_j(\lambda)$ carried out in the previous section. In fact, the steps to perform are the same, since one can write down a $Y$-system corresponding to the boundary partition function \eqref{eq:modified_partition_function} and solve it analytically for $\beta\to 0$. Since the calculations go through along the very same lines as before (although they are a bit more involved), they will not be repeated. We stress here that it is important to arrive at an exact solution only up to the first order in $s$, due to \eqref{eq:implicit_rho_j}. At the end of the computation, one obtains
\bea
\fl \eta_{1,s}^{(1)}(\lambda)=\frac{1}{|\kappa_{11}^2\kappa_{22}^2\kappa_{33}^2|}\frac{\omega^{+}(s,\lambda,|\kappa_{11}|,|\kappa_{22}|,|\kappa_{33}|)\omega^{-}(s,\lambda,|\kappa_{11}|,|\kappa_{22}|,|\kappa_{33}|)}{\phi_1(s,\lambda)\overline{\omega}(s,\lambda,|\kappa^{-1}_{11}|,|\kappa^{-1}_{22}|,|\kappa^{-1}_{33}|)}-1\,,\label{eq:analytical_eta1_s}\\
\fl \eta_{1,s}^{(2)}(\lambda)=|\kappa_{11}^2\kappa_{22}^2\kappa_{33}^2|\frac{\overline{\omega}^{+}(s,\lambda,|\kappa^{-1}_{11}|,|\kappa^{-1}_{22}|,|\kappa^{-1}_{33}|)\overline{\omega}^{-}(s,\lambda,|\kappa^{-1}_{11}|,|\kappa^{-1}_{22}|,|\kappa^{-1}_{33}|)}{\phi_2(s,\lambda)\omega(s,\lambda,|\kappa_{11}|,|\kappa_{22}|,|\kappa_{33}|)}-1\,.\label{eq:analytical_eta2_s}
\eea
Here we defined $\omega^{\pm}(s,\lambda,a,b,c)=\omega(s,\lambda\pm i/2,a,b,c)$ and
\bea
\fl \omega(s,u,a,b,c)= u^4(a^2+b^2+c^2)+2 u^2 (a^2+b^2+c^2)+a^2+b^2+c^2\nonumber\\
+\frac{4 i s \left[(u^2+1)(a^4+b^4+c^4)+a^2b^2+a^2c^2+b^2c^2\right]}{a^2+b^2+c^2}\,,\\
\fl \overline{\omega}(s,u,a,b,c)=\frac{1}{16}(a^2 + b^2 + c^2) (9 + 4 u^2)^2\nonumber\\
\fl  +\frac{
3 i s [(9+4u^2)(a^4b^2+a^2b^4+a^4c^2+a^2c^4+b^4c^2+b^2c^4)+(17+4u^2)a^2b^2c^2]}{
2 (a^2 b^2 + a^2 c^2 + b^2 c^2)}\,.
\eea
and
\bea
\phi_1(s,u)&=&\frac{u^2 \left(u^2-\left(s+\frac{i}{2}\right)^2\right)^2}{u^2+\frac{1}{4}}\,,\\
\phi_2(s,u)&=&\frac{u^2 \left(u^2-(-s+2 i)^2\right)^2}{u^2+\frac{1}{4}}\,.
\eea
Higher functions are obtained recursively from the formula
\bea
\fl \eta_{n,s}^{(1)}(\lambda,s)=\frac{\eta^{(1)}_{n-1,,s}(\lambda+i/2,s)\eta^{(1)}_{n-1,s}(\lambda-i/2,s)\left[1+\left(\eta_{n-1,s}^{(2)}\right)^{-1}\right]}{1+\eta_{n-2,s}^{(1)}(\lambda,s)}-1\,,\\
\fl \eta_{n,s}^{(2)}(\lambda,s)=\frac{\eta_{n-1,s}^{(2)}(\lambda+i/2,s)\eta^{(2)}_{n-1,s}(\lambda-i/2,s)\left[1+\left(\eta_{n-1,s}^{(1)}\right)^{-1}\right]}{1+\eta_{n-2,s}^{(2)}(\lambda,s)}-1\,,
\eea
with the convention $\eta_{0,s}^{(1)}(\lambda)=\eta_{0,s}^{(2)}(\lambda)=0$. Finally, one can work out explicitly the derivative in \eqref{eq:implicit_rho_j}, even though in general the ensuing analytic expression will be very complicated. As an example, the rapidity distribution functions $\rho^{(r)}_{t,1}(u)$ corresponding to the delta-state \eqref{eq:deltastate} read
\bea
\rho^{(1)}_{t,1}(u)=\frac{1}{\pi} \frac{8(80u^4+168u^2+53)}{(4u^2+1)(8u^2+3)(4u^2+9)^2},\\
\rho_1^{(2)}(u)=\frac{1}{2\pi} \frac{(4u^2+1)(5u^4+18u^2+8)}{(u^2+1)^2(u^2+4)^2(8u^2+3)}.
\eea
The expression for higher functions $\rho^{(r)}_{t,j}(u)$ is increasingly more complicated as $j$ increases.

\begin{figure}
	\centering
	\includegraphics[scale=0.75]{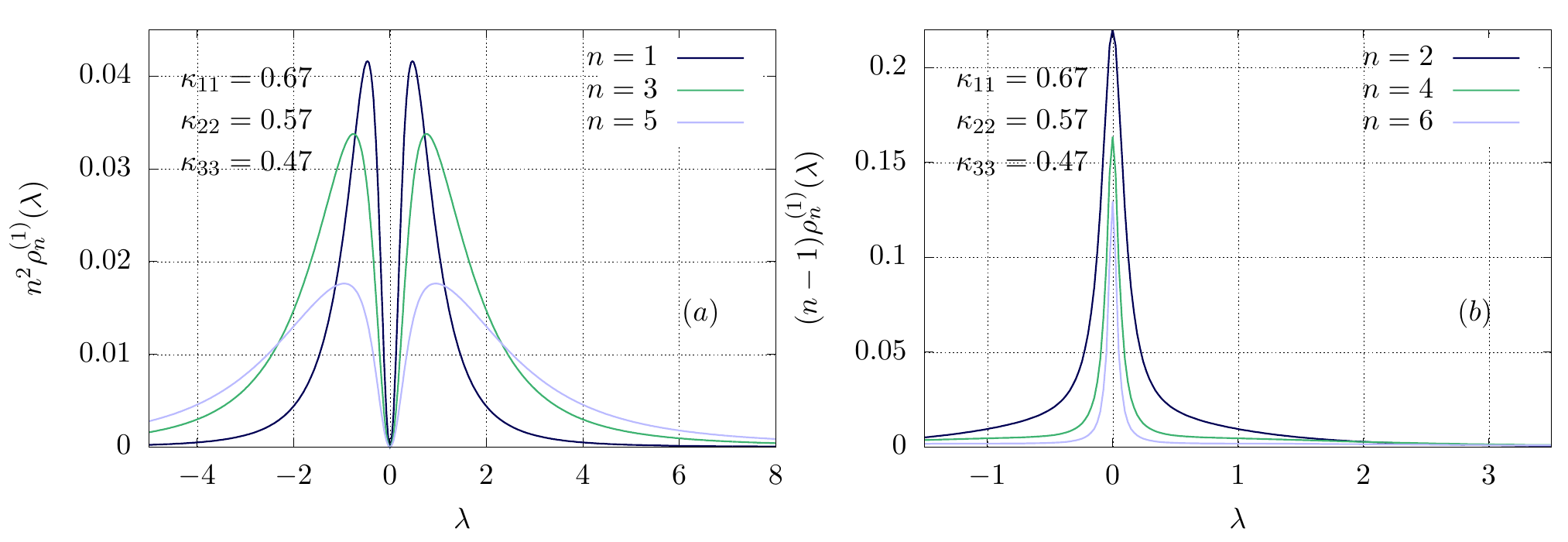}
	\caption{Rapidity distribution functions of the post-quench steady state for the quench from the state \eqref{eq:product_states} with \eqref{eq:integrable_block_diagonal} and $\kappa_{11}\simeq 0.67$, $\kappa_{22}\simeq 0.57$, $\kappa_{33}\simeq 0.47$. The plot shows the distributions corresponding to the first species of quasiparticles. Note the different rescaling employed for odd [subfigure $(a)$] and even [subfigure $(b)$] distributions.}
	\label{fig:roots1}
\end{figure}

\begin{figure}
	\centering
	\includegraphics[scale=0.75]{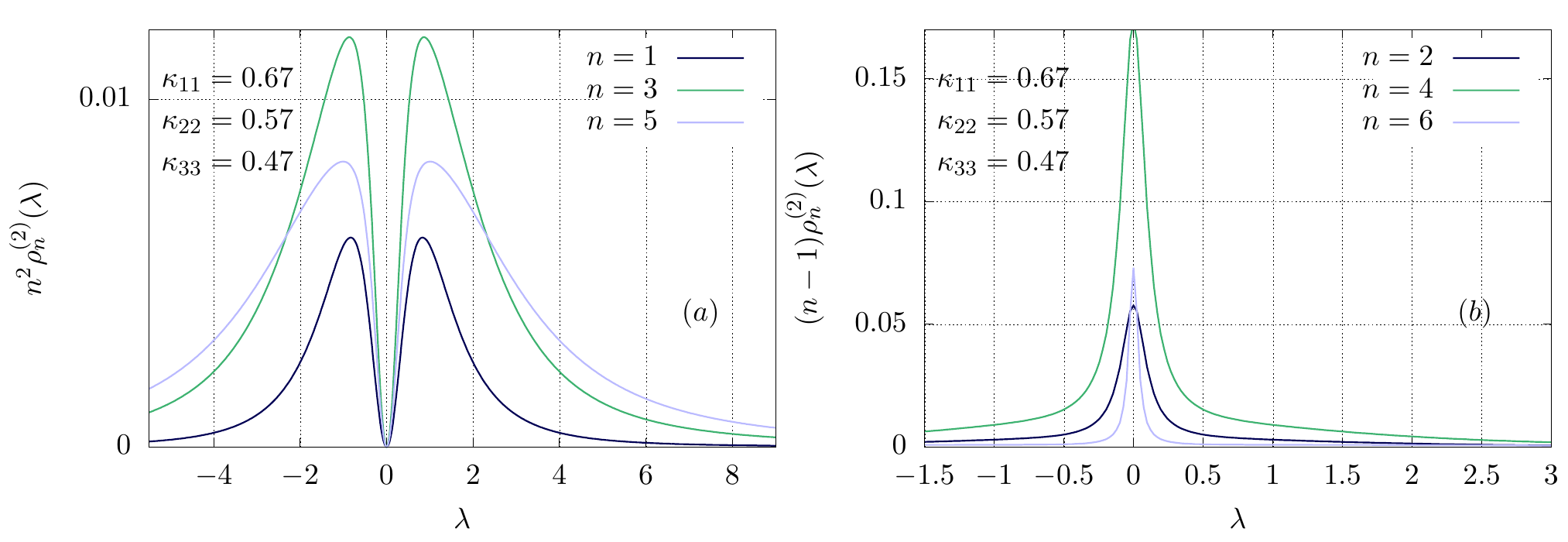}
	\caption{Rapidity distribution functions of the post-quench steady state for the quench from the state \eqref{eq:product_states} with \eqref{eq:integrable_block_diagonal} and $\kappa_{11}\simeq 0.67$, $\kappa_{22}\simeq 0.57$, $\kappa_{33}\simeq 0.47$. The plot shows the distributions corresponding to the second species of quasiparticles. Note the different rescaling employed for odd [subfigure $(a)$] and even [subfigure $(b)$] distributions.}
	\label{fig:roots2}
\end{figure}

As a consistency check, we explicitly verified that the numerical solution to \eqref{eq:TBAexplicit1} and \eqref{eq:TBAexplicit2} (using the functions $\eta^{(r)}_j(\lambda)$ previously derived) coincides, up to the expected numerical error, with the analytic result found above. We plot our results for the distribution functions for a given initial state in Figs.~\ref{fig:roots1} and \ref{fig:roots2}. We see that they display qualitatively different properties with respect to those of the initial state studied in \cite{MBPC17}: indeed, for $n$ odd they are vanishing at zero, and exhibit a non-monotonic behavior in $\lambda$ for $\lambda>0$. Furthermore, we see that the rapidity distribution function associated with $2$-particle bound states is dominant with respect to the unbound particles, and this is true for both species. This is once again in contrast to the case studied in \cite{MBPC17}, and also from the thermal case. Physically, this can be heuristically understood looking at the initial state, which is made of two-site blocks where the magnonic excitations always come in pairs. Indeed, the post-quench rapidity distribution function is determined by the states overlapping the most with the initial state. And for the diagonal two-site block \eqref{eq:integrable_block_diagonal}, we expect that it has large overlap with states containing many $2$-strings, which can be thought of as bound states of two magnons, spatially close to one another.

\begin{figure}
	\centering
	\includegraphics[scale=0.75]{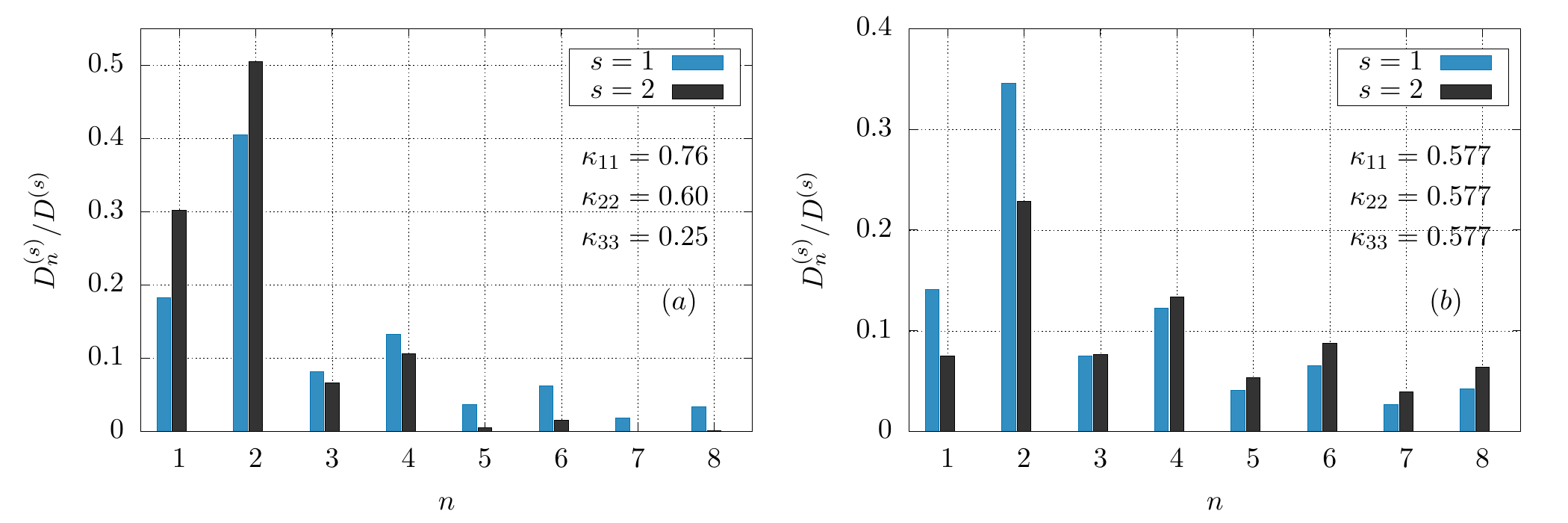}
	\caption{Composition in terms of bound states of the post-quench stationary state for the first and second species. Subfigures $(a)$ and $(b)$ correspond to different choices of the initial parameters $\kappa_{ii}$ for the initial state \eqref{eq:product_states} with \eqref{eq:integrable_block_diagonal}. The plots show in particular the relative densities $D^{(r)}_{n}/D^{(s)}$, where $D_n^{(r)}$ is defined in \eqref{eq:boundstate_density}, while $D^{(s)}$ is the total densities of quasiparticles of species $s$.}
	\label{fig:densities}
\end{figure}

In order to analyze this point further, it is useful to look at the composition of the post-quench steady state in terms of bound states of quasiparticles. In particular, we define the density of bound states for both species as
\bea
D^{(r)}_n&=&n\int_{-\infty}^{+\infty}\!\!\!\!{\rm d}k\,\, \rho_{n}^{(r)}(k)\,.\label{eq:boundstate_density}
\eea
Their distribution is shown in Fig.~\ref{fig:densities} for different initial states. From the plots, it is apparent that $2$-particle bound states are dominant. Furthermore, in general, we see that bound states of an even number of quasiparticles are more common than those with an odd number of them.

We stress that besides giving qualitative information on the nature of the post-quench steady state, the rapidity distribution functions obtained in this section give us access, in principle, to all the thermodynamic properties of the steady state. In the case of $XXZ$ Heisenberg chains, for instance, exact formulas have been derived which express correlation functions on an arbitrary macro-state solely in terms of its rapidity distribution functions \cite{MePo14,Pozs17}. While such formulas have not been generalized yet to the $SU(3)$-invariant Hamiltonian, we believe that this goal is fully within the reach of current research, making the results of this section an important building block. In Sec.~\ref{Sec:entanglement} we will instead present an immediate physical application of our findings.

\subsection{Analytical checks}

In order to provide a non-trivial check on the validity of our results, we compute the local conservation laws using the post-quench rapidity distribution functions and verify that this gives us the same result as when they are computed on the initial state.

First, we define the generating function $\Omega_{\Psi_0}(\lambda)$ such that
\be
\frac{\partial^{n}}{\partial\lambda^{n}}\Omega_{\Psi_0}(\lambda)\Big|_{\lambda=0}=\lim_{L\rightarrow\infty}\frac{1}{L}\langle \Psi_0| Q_{n+2}|\Psi_0\rangle\,.
\label{eq:def_generating_function}
\ee
It was shown in \cite{FaEs13,FCEC14} that the latter can be computed as
\be
\Omega_{\Psi_0}(\lambda)=i\lim_{L\to\infty}\frac{1}{L}\langle\Psi_0|t^{\dagger}(\lambda)\partial_{\mu}\big|_{\mu=\lambda}t(\mu)|\Psi_0\rangle\,,\label{eq:computation_omega}
\ee
where $t(\lambda)$ is the periodic transfer matrix \eqref{eq:periodic_transfer_matrix} (see also \cite{MBPC17} for the case of the $SU(3)$-invariant  spin chain). The computation of \eqref{eq:computation_omega} can be easily performed for the product states \eqref{eq:product_states}, see \ref{sec:generating_function}. In fact, it is possible to write down an analytic expression for $\Omega_{\Psi_0}(\lambda)$ for arbitrary choices of $\kappa_{jj}$, albeit the latter is in fact extremely unwieldy. We do not report it here, even though it can be obtain straightforwardly following the calculations of \ref{sec:generating_function}. As an  exception, in the case $\kappa_{11}=\kappa_{22}=\kappa_{33}=1/\sqrt{3}$, the generating function is extremely simple, and reads
\be
\Omega_{\Psi_0}(\lambda)=-\frac{2 \lambda ^2+1}{3 \left(\lambda ^2+1\right)^2}\,,\qquad \kappa_{11}=\kappa_{22}=\kappa_{33}=\frac{1}{\sqrt{3}}\,.
\ee

Next, given a state characterized by the rapidity distribution functions $\rho^{(r)}_j(\lambda)$, we can use the following formula for the expectation value of the conserved charges of the Hamiltonian \eqref{eq:hamiltonian} \cite{MBPC17}
\be
\fl \lim_{L\rightarrow\infty}\frac{1}{L}\braket{|\{\rho^{(r)}_{n}\}|  Q_m|\{\rho^{(r)}_{n}\}}=\sum_{n=1}^{\infty}\int_{-\infty}^{+\infty}{\rm d}\lambda\left( \rho^{(1)}_{n}(\lambda)c^{(1)}_{m,n}(\lambda)+\rho^{(2)}_{n}(\lambda)c^{(2)}_{m,n}(\lambda)\right)\,,
\label{eq:charge1}
\ee
where 
\bea
c^{(1)}_{m+1,n}(k)&=&(-1)^mi\frac{\partial^m}{\partial \lambda^m}\log\left[\frac{k+in/2}{k-in/2}\right]\,,\qquad\qquad m\geq 1\,,\label{eq:charge2}\\
c^{(2)}_{m+1,n}(\lambda)&\equiv& 0\,, \qquad\qquad m\geq 1\,.\label{eq:charge3}
\eea
We indicated with $\ket{\{\rho^{(r)}_{n}\}}$ a Bethe state which corresponds to the rapidity distributions $\{\rho^{(r)}_{n}\}$ in the thermodynamic limit.

We have checked (for values of $n$ up to $n=6$) that the expectation values of the charges $Q_n$ obtained from the generating function in \eqref{eq:computation_omega} coincide with the ones computed plugging in \eqref{eq:charge1} the post-quench rapidity distribution functions derived in the previous section, for several values of the initial-state parameters $\kappa_{jj}$. This provides the main test of our findings.

It is also interesting to compute the densities of the two quasiparticle species on the initial state:
\bea
D_1&=&\lim_{L\rightarrow\infty}\frac{1}{L}\braket{\Psi_0|\mathcal{N}_{1}|\Psi_0}=\frac{|\kappa_{22}|^2+|\kappa_{33}|^2}{|\kappa_{11}|^2+|\kappa_{22}|^2+|\kappa_{33}|^2}\,, \label{eq:particle_number1}\\
D_2&=&\lim_{L\rightarrow\infty}\frac{1}{L}\braket{\Psi_0|\mathcal{N}_{2}|\Psi_0}=\frac{|\kappa_{33}|^2}{|\kappa_{11}|^2+|\kappa_{22}|^2+|\kappa_{33}|^2}\,,
\label{eq:particle_number2}
\eea
where
\bea
\mathcal{N}_1 &\equiv& \sum_{j=1}^L\Bigl[ (E_2^2)_j + (E_3^3)_j\Bigr]\,, \label{eq:n1}\\
\mathcal{N}_2 &\equiv& \sum_{j=1}^L \Bigl[ (E_3^3)_j\Bigr]\,,\label{eq:n2}
\eea
and $E_i^j$ was defined in \eqref{eq:eij_operator}.
We find that for the integrable initial states \eqref{eq:product_states} with diagonal building block \eqref{eq:integrable_block_diagonal} (and under the restriction \eqref{eq:maximal_magnetization}) the above densities coincide with those computed using the post-quench rapidity distribution functions, cf. Eqs.~\eqref{eq:density1} and \eqref{eq:density2}. As already anticipated, we conclude that for these quenches the density of rapidities at infinity for the post-quench steady state is vanishing, in analogy with the case of quenches from the N\'eel state in $SU(2)$-invariant spin chains \cite{DWBC14_II}.

\section{A conjecture: the exact overlap formulas }\label{sec:overlaps}

Having solved the problem of characterizing the post-quench
steady state, it is tempting to wonder whether exact overlap formulas
between the integrable states and the eigenstates of the system can be
found. We remind that this was possible in all earlier cases studied in the literature, including the integrable
states of the XXZ Heisenberg chain, the Lieb-Liniger model, and even the MPS of the
nested chains studied in the context of the AdS/CFT correspondence \cite{PiPV17_II}.

Motivated by these examples in the existing literature, in this
section we conjecture overlap formulas between the integrable states
and the eigenstates of the Hamiltonian. On the one hand, we tested these
formulas against numerical calculations performed via exact
diagonalization for small system sizes; on the other hand, we verified that the ensuing Quench
Action treatment provides the same result as  the QTM approach,
giving us strong evidence on the validity of our conjectures. This will be shown explicitly in the next subsections.

\subsection{The overlap at finite size}

As we already stressed, the most difficult step in the application of
the quench action method, is the computation of the overlaps between
the initial state and the eigenstates of the Hamiltonian. Indeed, no
general procedure has been devised yet for this task. 

In the prototypical case of the XXZ Heisenberg chain, this problem has
been essentially solved for integrable initial states
\cite{Pozs18}. In this case, the overlaps display a remarkably simple
``universal'' form: namely they are proportional to a ratio of
Gaudin-like determinants, which do not depend on the initial state,
and a state-dependent prefactor which is a product of single-particle
functions. A rigorous proof of the formulas of
\cite{Pozs18} is only available for states associated to the
diagonal $K$-matrices \cite{Pozs14,Broc14}; nevertheless the single-particle overlap
functions and some additional numerical pre-factors could be fixed
unambiguously by comparing the Quench Action and QTM computations for the
Loschmidt echo. The full structure of the overlap including the
Gaudin-like determinants was tested and confirmed numerically at
small system sizes.

A completely analogous form of the overlaps was already observed for
specific integrable states of the $SU(3)$- and $SO(6)$-invariant spin
chains in \cite{LeKM16,MBPC17,LeKL18}, where MPSs relevant
to the AdS/CFT conjecture were studied. These overlap formulas were
obtained from coordinate Bethe Ansatz for small particle numbers, and
also tested numerically for more complicated eigenstates. 
They involve the ratio of the Gaudin-like determinants associated to
the nested Bethe equations of the models, and a pre-factor which is a
linear combination of certain products of the one-particle overlap
functions. The MPSs studied in these works are outside the scope of
the present article, and are investigated in \cite{InPrep_II}. Nevertheless the aforementioned results serve as an
inspiration and they lead us to a conjecture for the overlaps of the integrable two-site product states studied in this work.

Let $|\Psi_0\rangle$ be a product state with the two-site building block given in \eqref{eq:integrable_block}. First, since the state is integrable, it has non-vanishing overlaps only with parity invariant eigenstates, namely with those corresponding to sets of rapidities
\bea
\{ k_j\}_{j=1}^{N}=\{ k_j^+\}_{j=1}^{N/2}\cup \{- k_j^+\}_{j=1}^{N/2}\,,\\
\{\lambda_j\}_{j=1}^{N}=\{ \lambda_j^+\}_{j=1}^{N/2}\cup \{- \lambda_j^+\}_{j=1}^{N/2}\,.
\eea
We indicate such states with $|\{\pm k_j^+\}_{j=1}^{N/2},\{\pm\lambda_l^+\}_{l=1}^{M/2}\rangle$. Then, we conjecture the following overlap formula
\bea
\fl \frac{|\langle\Psi_{0}|\{\pm k_j^+\}_{j=1}^{N/2}\{\pm\lambda_l^+\}_{l=1}^{M/2}\rangle|^2}
{\langle \{\pm k_j^+\}\{\pm\lambda_j^+\}| \{\pm k_j^+\}\{\pm\lambda_l^+\}\rangle}
=\Gamma(\{\kappa_{ij}\}) \prod_{j=1}^{N/2}  h(k^{+}_j)
\prod_{j=1}^{M/2}  h(\lambda^{+}_j)
\times \frac{\det_{N/2}G_{jk}^{+}}{\det_{N/2}G_{jk}^{-}}\,,
\label{eq:overlap}
\eea
where 
\begin{equation}
\label{usym}
h(\lambda)=
\frac{\lambda^2}{\lambda^2+1/4}\,,
\end{equation}
while $G_{\pm}$ are Gaudin-like matrices defined by
\bea
G_{\pm}=\left(
\begin{array}{cc}
	A_{\pm}& B_\pm\\
	B^{t}_{\pm}& C_{\pm}
\end{array}
\right),
\eea
where
\bea
\fl \left(A_{\pm}\right)_{r,s}=\delta_{rs}\left[L \mathcal{K}_{1}(k_r)-\sum_{l=1}^{N/2} \mathcal{K}^{+}_{2}(k_r, k_l)+\sum_{l=1}^{M/2} \mathcal{K}_1^{+}(k_r, \lambda_l)\right]+\mathcal{K}^{\pm}_{2}(k_r,k_s)\,,\\
\fl \left(B_{\pm}\right)_{r,s}=-\mathcal{K}^{\pm}_{1}(k_r,\lambda_s)\,,\\
\fl \left(C_{\pm}\right)_{r,s}=\delta_{rs}\left[-\sum_{l=1}^{M/2} \mathcal{K}^{+}_{2}(\lambda_r, \lambda_l)+\sum_{l=1}^{M/2} \mathcal{K}_1^{+}(\lambda_r, k_l)\right]+\mathcal{K}^{\pm}_{2}(\lambda_r,\lambda_s)\,,
\eea
with the additional definitions
\bea
\mathcal{K}_{\ell}(u)&=&\frac{\ell}{u^2+\ell^2/4}\,,\\
\mathcal{K}_{s}^{\pm}(u,w)&=&\mathcal{K}_{s}(u-w)\pm \mathcal{K}_{s}(u+w)\,,\quad s=1,2\,.
\eea
Finally, we introduced 
\bea
\label{gammaov}
\Gamma(\{\kappa_{ij}\})
&=&
\kappa_{11}^{L-2N+M}
(\kappa_{11}\kappa_{22}-\kappa_{12}^2)^{N-2M} (\det(\kappa_{ij}))^{M}\,.
\eea 
The numbers $N, M$ are restricted to $0\leq N \leq 2L/3$, $0 \leq M
\leq N/2$. This condition ensures that we work on the correct ``side
of the equator'' \cite{Baxt02}.

At present the formula \eqref{eq:overlap} is a conjecture, but its
status is similar to the overlaps in the XXZ chain: once we assume a
structure like \eqref{eq:overlap} involving the Gaudin-like
determinants, and a product over one-particle overlap functions, then
all these pre-factors can be fixed unambiguously. In the following we
sketch this derivation.

First we consider the overlaps with the delta-state \eqref{eq:deltastate}. In the case
of $N\ge 2$, $M=0$ the Bethe states occupy only the basis states
$\ket{1}$ and $\ket{2}$, and the eigenstates are identical to those
of the $SU(2)$-invariant model, the XXX chain. For this case, the results
of   \cite{Pozs18} can be used after performing the isotropic limit of the XXZ Heisenberg chain. Defining the delta-state of the XXX model as
\be
| \psi_{\delta,2} \rangle = |1,1\rangle + |2,2\rangle \,,
\label{eq:deltastate2}
\ee
a simple calculation gives the normalized overlap in the XXX model as
\begin{equation}
\frac{|\langle\Psi_{\delta,2}|\{\pm\lambda^+\}_{N/2}\rangle|^2}
{\langle\{\pm\lambda^+\}_{N/2}|\{\pm\lambda^+\}_{N/2}\rangle}
=
\frac{1}{2^{L/2}}
\prod_{j=1}^{N/2}  h(\lambda_j)
\times \frac{\det_{N/2}G_{jk}^{+}}{\det_{N/2}G_{jk}^{-}}\,,
\label{OVERLAPSlekjopp2}
\end{equation}
with $h(\lambda)$ given by \eqref{usym}. Assuming that in the nested
case we have the same product structure, we are led to the formula
\bea
\fl \frac{|\langle\Psi_{\delta}|\{\pm k_j^+\}_{j=1}^{N/2}\{\pm\lambda_l^+\}_{l=1}^{M/2}\rangle|^2}
{\langle \{\pm k_j^+\}\{\pm\lambda_j^+\}| \{\pm k_j^+\}\{\pm\lambda_l^+\}\rangle}
=
\frac{1}{3^{L/2}}
\prod_{j=1}^{N/2}  h(k^{+}_j)
\prod_{j=1}^{M/2}  \tilde h(\lambda^{+}_j)
\times \frac{\det_{N/2}G_{jk}^{+}}{\det_{N/2}G_{jk}^{-}}\,,
\label{eq:overlappp}
\eea
where the difference in the numerical pre-factor (the normalization
with $3^{-L/2}$) comes from a difference between the norms of
$\ket{\Psi_\delta}$ and $\ket{\Psi_{\delta,2}}$, and $\tilde
h(\lambda)$ can be a new function independent from
$h(\lambda)$. However, when turning to the quenches in the
thermodynamic limit, the exact solution to the Quench Action equations displays the
symmetry $\eta^{(1)}_j(u)=\eta^{(2)}_j(u)$, cf. Eq.~\eqref{eq:analytical_etadelta} . In turn, this implies $\tilde h(\lambda)=h(\lambda)$.

Turning to the general diagonal states of the form
\eqref{eq:integrable_block_diagonal} simple coordinate Bethe Ansatz
arguments lead to
\bea
\frac{|\langle\Psi_{0}|\{\pm k_j^+\}_{j=1}^{N/2}\{\pm\lambda_l^+\}_{l=1}^{M/2}\rangle|^2}
{\langle \{\pm k_j^+\}\{\pm\lambda_j^+\}| \{\pm k_j^+\}\{\pm\lambda_l^+\}\rangle}
=
\frac{\kappa_{11}^{L-N}\kappa_{22}^{N-M}\kappa_{33}^{M}}{(|\kappa_{11}|^2+|\kappa_{22}|^2+|\kappa_{33}|^2)^{L/2}}\nonumber\\ 
\times 
\prod_{j=1}^{N/2}  h(k^{+}_j)
\prod_{j=1}^{M/2}   h(\lambda^{+}_j)
\times \frac{\det_{N/2}G_{jk}^{+}}{\det_{N/2}G_{jk}^{-}}\,,
\label{eq:overlapppp}
\eea
Finally, in the case of a general symmetric two-site block the pre-factor, Eq.~\eqref{gammaov} can be derived using group theoretical arguments,
building on the fact that the Bethe states are highest weight with
respect to the standard Cartan generators. 

We tested Eq.~\eqref{eq:overlap} extensively against exact
diagonalization calculations, using the following procedure. First, we found numerically different solutions to the Bethe equations \eqref{eq:bethe_equations1} and \eqref{eq:bethe_equations2}, corresponding to several eigenstates. By comparing the eigenvalues \eqref{eq:energy_eigenvalue} with the energy levels obtained by exact diagonalization of the Hamiltonian, the associated eigenstates were identified; such an identification was confirmed by computing the expectation values of higher-order conserved charges. The overlaps were then computed numerically, showing agreement (up to numerical precision) with the conjecture \eqref{eq:overlap}. It is worth noting at this point that our conjecture only specifies the squared norm of the overlaps rather than the overlaps themselves.
As it will be seen in the next section, an additional check of our conjecture comes from the application of the Quench Action approach using these overlaps, which leads to the correct QTM predictions.

\subsection{Comparison with the QTM results}

Assuming the validity of the conjecture presented in the previous section, we can compare our QTM predictions with the ones obtained by using the Quench Action approach, following the treatment of Sec.~\ref{sec:QAM}. We focus in particular on the diagonal case \eqref{eq:integrable_block_diagonal}. Furthermore, in this section we enforce the normalization condition
\be
\sum_{j=1}^{3}|\kappa_{jj}|^{2}=1\,.
\ee 
The first step in the application of the Quench Action method is the computation of the thermodynamically leading term of the overlaps \eqref{eq:overlap}, cf.~\eqref{eq:overlap_SQ}. In order to do so, we proceed as in \cite{MBPC17} to argue that the ratio of the determinants can be neglected, as its contributions are $O(1)$. It is straightforward to take the thermodynamic limit, which reads
\bea
\fl S_{\Psi_0}\left[\{\rho^{(r)}_n\}_{n=1}^{\infty}\right] =-\log(|\kappa_{11}|) -\frac{1}{4}\sum_{n=1}^{\infty}\int_{-\infty}^{\infty}{\rm d}k \rho^{(1)}_{n}(k) g_{n}(k)-\frac{1}{4}\sum_{n=1}^{\infty}\int_{-\infty}^{\infty}{\rm d}\lambda \rho^{(2)}_{n}(\lambda)g_{n}(\lambda)\nonumber\\
\fl +\frac{1}{2}\sum_{n=1}^{\infty}n\int_{-\infty}^{\infty}{\rm d}\lambda \rho^{(1)}_{n}(\lambda) \log\left(\frac{|\kappa_{11}|}{|\kappa_{22}|} \right)
+\frac{1}{2}\sum_{n=1}^{\infty}n\int_{-\infty}^{\infty}{\rm d}\lambda \rho^{(2)}_{n}(\lambda)\log\left(\frac{|\kappa_{22}|}{|\kappa_{33}|} \right)
\,,
\label{eq:sq_expression}
\eea
where
\bea
g_n(\lambda) &=& \sum_{l=0}^{n-1} \Big[ f_{n-1-2l}(\lambda) - f_{n-2l}(\lambda) \Big]\,, \label{eq:g_function}
\eea
and
\bea
f_n (\lambda) &=& \ln \big(\lambda^2 + n^2/4 \big) \label{eq:f_function}\,.
\eea

\begin{figure}
	\centering
	\includegraphics[scale=0.75]{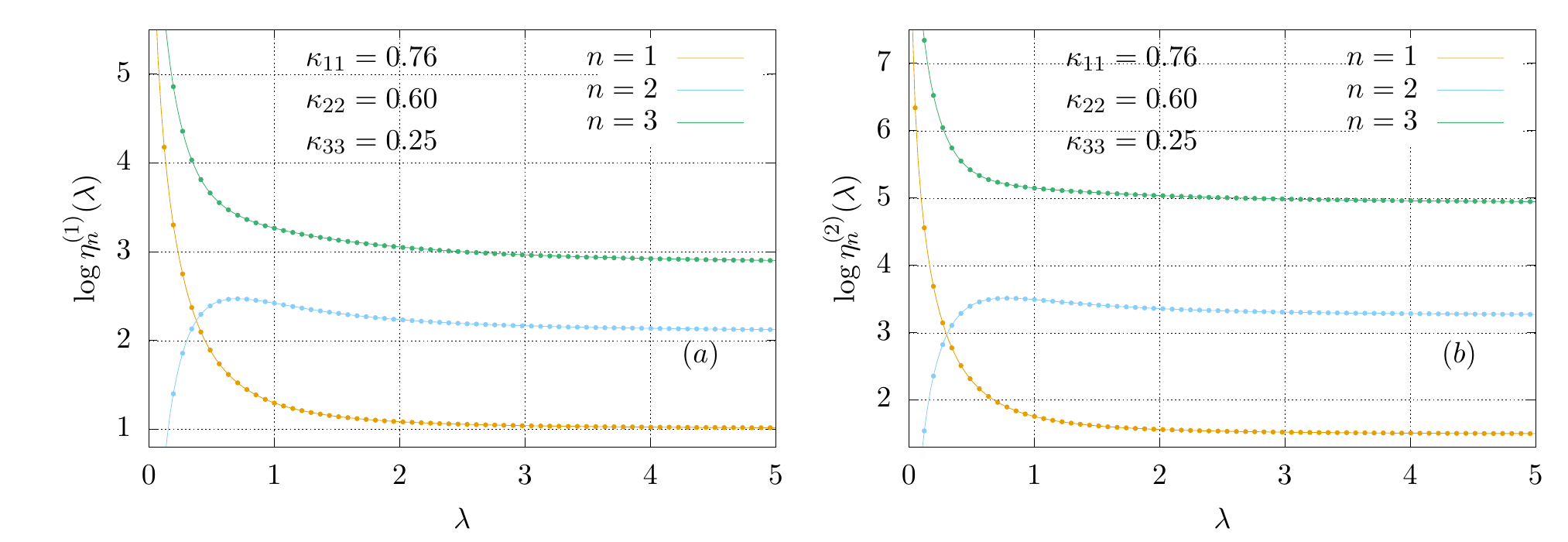}
	\caption{Distribution functions $\eta_j^{(r)}(\lambda)$ of the post-quench steady state for the quench from the state \eqref{eq:product_states} with \eqref{eq:integrable_block_diagonal} and $\kappa_{11}\simeq 0.76$, $\kappa_{22}\simeq 0.6$, $\kappa_{33}\simeq 0.25$. The plot shows the distributions corresponding to the first [subfigure $(a)$] and second [subfigure $(b)$] species of quasi-particles. Solid lines correspond to the analytical results, while dots are obtained via a numerical solution to the Quench Action equations.}
	\label{fig:comparison_etas}
\end{figure}

We can now perform explicitly the prescriptions of the Quench Action approach. In particular, taking the functional derivative in \eqref{eq:saddle_point} and exploiting the Bethe equations \eqref{eq:TBAexplicit1} and \eqref{eq:TBAexplicit2}, we obtain
\bea
\ln \eta_{n}^{(1)}(\lambda)&=&-g_n(\lambda)+\sum_{m=1}^{+\infty}\left[a_{n,m}\ast \ln\left(1+\left[\eta_m^{(1)}\right]^{-1}\right)\right](\lambda)\nonumber\\
&-&\sum_{m=1}^{+\infty}\left[b_{n,m}\ast \ln\left(1+\left[\eta_m^{(2)}\right]^{-1}\right)\right](\lambda)+2n\nu_1\,,\label{eq:eta_(1)}
\eea
\bea
\ln \eta_{n}^{(2)}(\lambda)&=&-g_n(\lambda)+\sum_{m=1}^{+\infty}\left[a_{n,m}\ast \ln\left(1+\left[\eta_m^{(2)}\right]^{-1}\right)\right](\lambda)\nonumber\\
&-&\sum_{m=1}^{+\infty}\left[b_{n,m}\ast \ln\left(1+\left[\eta_m^{(1)}\right]^{-1}\right)\right](\lambda)+2n\nu_2\,,\label{eq:eta_(2)}
\eea
where the functions $a_{n,m}$ and $b_{n,m}$ are defined in \eqref{eq:a_mn} and \eqref{eq:b_mn} respectively, while
\be
\nu_1=\log\frac{|\kappa_{11}|}{|\kappa_{22}|}\,, \qquad 
\nu_2=\log\frac{|\kappa_{22}|}{|\kappa_{33}|}\,.
\ee
These equations can be easily solved numerically by iteration, and their solution can be exploited to solve the thermodynamic form of the Bethe equations \eqref{eq:TBAexplicit1} and \eqref{eq:TBAexplicit2}, once again by numerical iteration. We compared systematically these results with the analytic formulas obtained using the QTM approach, finding always perfect agreement (see Fig.~\ref{fig:comparison_etas}, where an example is displayed). This provides strong evidence that the overlap formulas conjectured in the previous section are indeed correct. 

\section{A physical application: the entanglement dynamics}
\label{Sec:entanglement}

In this section, we finally provide a simple physical application of the results obtained so far. As we have anticipated, once the rapidity distribution functions of a given state are known, one can in principle compute all its thermodynamic properties. For instance, in the spin-$1/2$ Heisenberg chain, formulas have been derived \cite{MePo14,Pozs17} to compute correlations based solely on the knowledge of the rapidity distribution functions. Unfortunately, such formulas have not been derived yet for the $SU(3)$-invariant spin chains. However, there is an important physical quantity that we can already compute based on the knowledge of the post-quench rapidity distribution functions, namely the entanglement entropy. More precisely, in this section, we provide an exact prediction for the time evolution of the latter after the quench from an integrable initial state. 

We focus on the time-dependent entanglement entropy between an interval $A$ and the rest of the system ($\bar{A}$), defined as
\be
S_A(t)=-\textrm{tr}_A[\rho_A(t)\log\rho_A(t)]\,,
\label{eq:def_entanglement_entropy}
\ee
where $\rho_A(t)$ is the reduced density matrix corresponding to the subsystem $A$, namely
\be
\rho_A\equiv \lim_{L\to\infty}{\rm tr}_{\bar A}\left[e^{-iH_{L}t}|\Psi_0\rangle\langle \Psi_0|e^{iH_{L}t}\right]\,.
\ee 
The ab-initio computation of \eqref{eq:def_entanglement_entropy} in the presence of interactions is a highly non-trivial task, especially for large subsystem sizes $\ell$, 
when all the correlators within the subsystem need to be computed. 
However, recently these difficulties have been bypassed: in Ref. \cite{AlCa17} it has been argued that a quasiparticle argument provides a 
prediction for the entire time-evolution of the entanglement entropy which becomes exact  in the limit of large times $t$ and subsystem sizes $\ell$, 
but with their ratio arbitrary.
The approach of \cite{AlCa17} can be applied only when uncorrelated pairs of  quasiparticles are produced after the quench, which is precisely the case for integrable initial states \cite{PiPV17_II} (see Ref. \cite{BeTC17} for a generalization of this result  to the case of quenches which produce $n$-tuplets of quasiparticles   and Ref. \cite{bc-18}
for the case of correlated pairs, both valid only for free systems). 
We mention that these arguments may be adapted to inhomogeneous situations as well \cite{transport,transport2} by exploiting the  recently developed 
generalized hydrodynamics \cite{ghd}.

The results of \cite{AlCa17} were initially applied to quenches in the $XXZ$ Heisenberg chain, and have been extensively tested against  numerical simulations based 
on time-dependent density matrix renormalization group \cite{white-2004,daley-2004,uli-2011} and infinite time evolved block decimation \cite{Vida07} algorithms. 
The arguments underlying their derivation are very general and can be straightforwardly followed also in the case of nested integrable models. 
In fact, a generalization of the formulas of \cite{AlCa17} was already presented in \cite{MBPC17}, where predictions for the entanglement dynamics were provided for the initial state \eqref{eq:initial_state}. In particular, in the limit of large $\ell$ and $t$, one has the following expression for the entanglement dynamics 
\be
\fl S_\ell(t)=  \sum_{r=1,2}\,\sum_{n=1}^{\infty}\,\int\!\!{\rm d}\lambda\,\, s_{n}^{(r)}(\lambda)\left\{2 t |v_{n}^{(r)}(\lambda)|\,\theta_{\rm H}(\ell-{2|v_{n}^{(r)}(\lambda)|t})+\ell\, \theta_{\rm H}({2|v_{n}^{(r)}(\lambda)|t}-\ell)\right\}\,.
\label{eq:entanglemententropy}
\ee
Here $\theta_{\rm H}(x)$ is the Heaviside step function, while the Yang-Yang entropy density $s_{n}^{(r)}(\lambda)$ is given by  
\bea
 s_{n}^{(r)}(\lambda)&=&\left(\rho_n^{(r)}(\lambda)+\rho_{h,n}^{(r)}(\lambda)\right)\ln\left(\rho_n^{(r)}(\lambda)+\rho_{h,n}^{(r)}(\lambda)\right)\nonumber\\
&-&\rho_{n}^{(r)}(\lambda)\ln\rho_{n}^{(r)}(\lambda)-\rho_{h,n}^{(r)}(\lambda)\ln\rho_{h,n}^{(r)}(\lambda)\,.
\label{eq:yang_yang}
\eea
Finally, the velocities $v_n^{(1)}(\lambda)$ and $v_n^{(2)}(\lambda)$ fulfill the integral equations
\bea
\fl \rho^{(2)}_{t,n}(\lambda) v^{(2)}_n(\lambda)=\sum_{k}\left(b_{n,k}\ast v^{(1)}_k \rho^{(1)}_k\right)(\lambda)-\sum_{k}\left(a_{n,k}\ast v^{(2)}_k \rho^{(2)}_k\right)(\lambda),\label{eq:velocities1}\\
\fl \rho^{(1)}_{t,n}(\lambda)v^{(1)}_n(\lambda)=\frac{1}{2\pi}\varepsilon^{\prime}_n(\lambda)-\sum_k \left( a_{n,k}\ast v^{(1)}_k\rho_k^{(1)}\right)(\lambda)+\sum_k \left(b_{n,k} \ast v^{(2)}_k \rho_k^{(2)}\right)(\lambda)\,,
\label{eq:velocities2}
\eea
where $\varepsilon_n(\lambda)$ is defined in \eqref{eq:epsilon_energy}. Note that a fundamental input of these formulas is the knowledge of the quasiparticle rapidity distribution functions $\rho_{n}^{(r)}(\lambda)$ of the post-quench steady state. We refer the reader to \cite{AlCa17, MBPC17} for a thorough discussion of these formulas, while here we simply sketch the main ideas underlying their heuristic derivation. 
A mathematical proof of Eq. \eqref{eq:entanglemententropy} is so far available only for free fermionic theories \cite{fagotti-2008}.

The standard quasiparticle picture, introduced in \cite{CaCa05}, works as follows. 
The quench is interpreted as a process generating everywhere and homogeneously uncorrelated pairs of entangled quasiparticles of opposite momenta, 
hence moving ballistically in opposite directions. 
This assumption is very reasonable for integrable initial states. 
Since the quasiparticles emitted from different points are unentangled, two regions may be entangled only if there is at least a pair
of quasiparticles shared between them, emitted from an arbitrary common point.
Hence the total entanglement entropy between a region $A$ and the rest of the system is related to number of pairs 
with one quasiparticle in $A$ and the other in its complement $\bar A$. 
Two additional ingredients are needed in order to derive \eqref{eq:entanglemententropy}. 
First, one needs to determine the velocities $v^{(r)}_n(\lambda)$ of the quasiparticles: these are obtained as the group velocities of the
elementary excitations over the stationary state corresponding to $\rho^{(r)}_n(\lambda)$, which can be shown to fulfill the set of integral equations \eqref{eq:velocities1} and \eqref{eq:velocities2}, as shown in \cite{BoEL14}. 
Second, one needs to work out the contribution to the entanglement carried by each quasiparticle pair.
Given that the stationary value of  the entanglement entropy of a large subsystem has the same density as the thermodynamic entropy of the stationary state \cite{AlCa17}, one
naturally identifies such a contribution with the Yang-Yang entropy density \eqref{eq:yang_yang}. 
With these ingredients, and following a simple heuristic derivation, Eq.~\eqref{eq:entanglemententropy} is easily derived \cite{AlCa17,MBPC17}.

\begin{figure}
	\centering
	\includegraphics[scale=0.75]{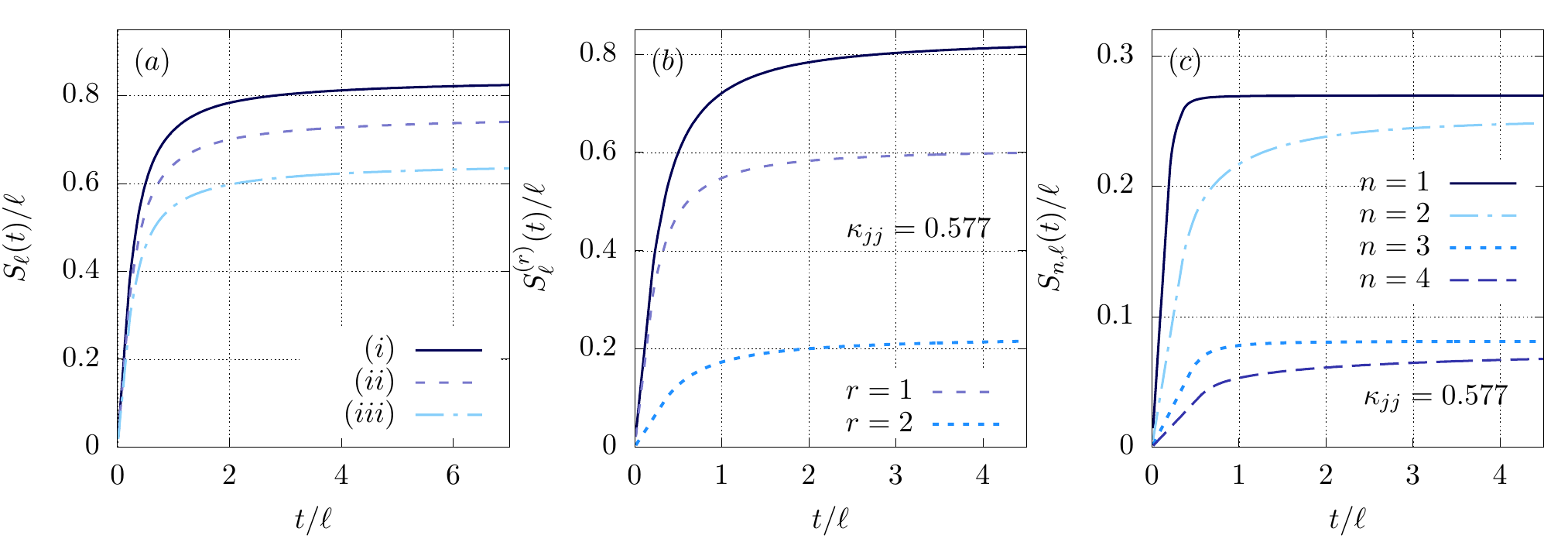}
	\caption{Time-dependent entanglement entropy \eqref{eq:entanglemententropy} after a quench from integrable initial states \eqref{eq:product_states} with the diagonal block $\eqref{eq:integrable_block_diagonal}$. Subfigure $(a)$: the time evolution of the entanglement entropy as a function of $t/\ell$. Different curves correspond to different parameters of the initial states. $(i): \kappa_{11}=\kappa_{22}=\kappa_{33}=1/\sqrt{3}\simeq 0.577$; $(ii):\kappa_{11}=0.7, \kappa_{22}=0.6, \kappa_{33}=0.387$; $(iii):\kappa_{11}=0.8, \kappa_{22}=0.5, \kappa_{33}=0.332$. Subfigure $(b)$: contribution of the different quasiparticle species to the entanglement dynamics. The initial state is as for the case $(i)$ of subfigure $(a)$. Dashed lines correspond to the contribution of the two species, while the solid line indicates their sum, namely the full entanglement $S_{\ell}(t)/\ell$. Subfigure $(c)$: contribution of the different $n$-quasiparticle bound states to the entanglement dynamics. The initial state is as for the case $(i)$ of subfigure $(a)$.}
	\label{fig:entanglement}
\end{figure}

In Fig.~\ref{fig:entanglement}, we report the time evolution of the entanglement entropy after a quench from integrable initial states \eqref{eq:product_states}, with different choices of the diagonal block \eqref{eq:integrable_block_diagonal}. The qualitative behavior of the plots is the one expected from the quasiparticle picture of \cite{CaCa05}, and already observed in \cite{MBPC17} for the initial state \eqref{eq:initial_state}. For $t< \ell/2v_{\rm max}$ the growth of entanglement is linear and governed by the fastest quasiparticles; after this time, the entanglement displays a saturation, where all the slower quasiparticles start to contribute. Differently to \cite{MBPC17}, we have the freedom to tune the parameters $\kappa_{jj}$ of the initial-state building block \eqref{eq:integrable_block_diagonal}. We see from subfigure $(a)$ of Fig.~\ref{fig:entanglement} a strong quantitative dependence on the initial parameters.

In subfigures $(b)$ and $(c)$ we plot the contribution to the entanglement coming from different species
\be
\fl S^{(r)}_\ell(t)= \sum_{n=1}^{\infty}\,\int\!\!{\rm d}\lambda\,\, s_{n}^{(r)}(\lambda)\left\{2 t |v_{n}^{(r)}(\lambda)|\,\theta_{\rm H}(\ell-{2|v_{n}^{(r)}(\lambda)|t})+\ell\, \theta_{\rm H}({2|v_{n}^{(r)}(\lambda)|t}-\ell)\right\}\,,
\label{eq:entanglemententropy_per_species}
\ee 
as well as the contribution of the different bound-states of quasiparticles
\be
\fl S_{n,\ell}(t)= \sum_{r=1,2}\,\int\!\!{\rm d}\lambda\,\, s_{n}^{(r)}(\lambda)\left\{2 t |v_{n}^{(r)}(\lambda)|\,\theta_{\rm H}(\ell-{2|v_{n}^{(r)}(\lambda)|t})+\ell\, \theta_{\rm H}({2|v_{n}^{(r)}(\lambda)|t}-\ell)\right\}\,,
\label{eq:entanglemententropy_per_boundstate}
\ee 
for the delta-state (namely choosing the initial building block \eqref{eq:deltastate}). We see that the first species has a larger contribution; furthermore, while it is true that unbound particles carry the largest amount of entanglement, the one of the two-quasiparticle bound states is comparable. Note that this is expected, since their density is dominant, cf. Fig.~\ref{fig:densities}; furthermore, this is in contrast to what was found in \cite{MBPC17} for the state \eqref{eq:initial_state}.

Following \cite{MBPC17} it is interesting to also compute the evolution of the mutual information between two disjoint spin blocks $A$ and $B$ of length $\ell$, separated by a distance $d$
\be
I_{A:B}=S_A+S_B-S_{A\cup B}\,.
\label{eq:def_mutual_info}
\ee
Indeed, this quantity is better suited to resolve the contribution of different species and bound states. Within the quasiparticle picture, the mutual information can be computed by counting all the pairs of quasiparticles with one quasiparticle in $A$ and the other in 
$B$, leading to the formula \cite{MBPC17}
\bea
I_{A:B}(t)&=&\sum_{r=1,2}\sum_{n=1}^\infty \int\!\!\mathrm{d}\lambda\,\biggl[\left(2|v^{(r)}_n(\lambda)| t - d\right)\chi_{[d,d +\ell]}(2 |v^{(r)}_n(\lambda)| t)\nonumber\\
&+& \left(d + 2 \ell - 2 |v^{(r)}_n(\lambda)| t\right)\chi_{[d+\ell,d +2\ell]}(2 |v_n^{(r)}(\lambda)| t)\biggr]s_n^{(r)}(\lambda)\,,
\label{eq:mutual_info_def}
\eea
where $\chi_{[a,b]}(x)$ is the characteristic function of $[a,b]$, \emph{i.e.} it is equal to $1$ if $x\in[a,b]$ and equal to $0$ otherwise. 

\begin{figure}
	\centering
	\includegraphics[scale=0.75]{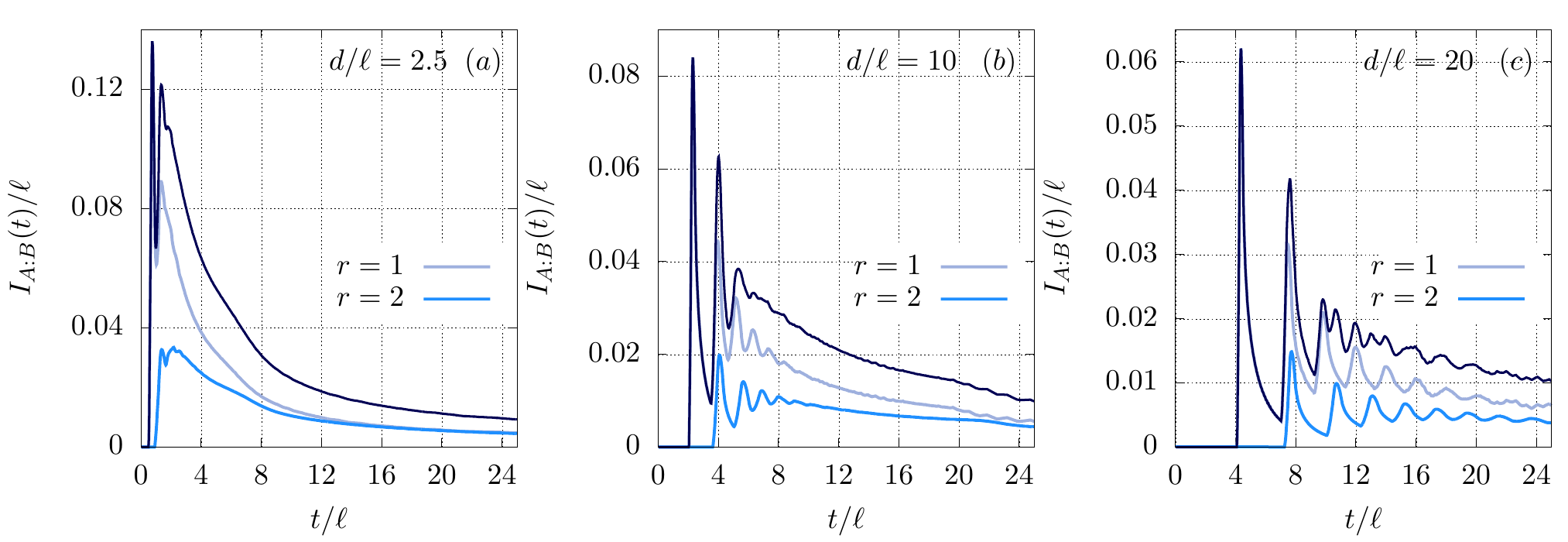}
	\caption{Time-dependent mutual information \eqref{eq:mutual_info_def} after a quench from the integrable initial state \eqref{eq:product_states} with diagonal block \eqref{eq:deltastate} (delta-state). Subfigure $(a)$, $(b)$ and $(c)$ correspond to values of the ratio $d/\ell= 2.5, 10$, and  $20$ respectively. In each plot we report the total mutual information (blue solid
		line) together with the separate contributions carried by each species of quasiparticles. As $n$ increases, the velocity of the $n$-quasiparticle bound states decreases: hence, subsequent peaks correspond to $n$-quasiparticle bound states with increasing values of~$n$.}
	\label{fig:mutual_info}
\end{figure}

The time evolution of the mutual information is reported in Fig.~\ref{fig:mutual_info}, where we plot $I_{A:B}(t)/\ell$ against $t/\ell$ for the values of ratio $d/\ell=2.5, 10$ and $20$. Peaks are clearly visible, signaling the contribution of the different types of quasiparticles (namely different species and bound states). As the ratio $d/\ell$ increases the peaks separate, so that the different contribution of the quasiparticles become more visible. 

Two other important quantities related to the entanglement entropy can be also computed from the results presented in this paper. 
These are the R\'enyi entanglement entropies $[{\rm tr} (\rho_A^n)]/(1-n)$ and the entanglement negativity between two blocks of spins. 
These quantities may be obtained, following the  approach of Refs. \cite{AlCa17a,AlCa17b,mestyan-2018,ac-18n}, 
only in the stationary state and it is still an open problem to understand their entire time evolution.
However, also the calculation of the stationary values is rather cumbersome since it requires the numerical solution of some integral equations depending explicitly on the overlaps \cite{mestyan-2018}.
For these reasons, we postpone their computation to future investigations.

\section{Conclusions}
\label{concl}

In this work we have applied the recently introduced QTM approach to
quantum quenches in the prototypical $SU(3)$-invariant spin chain \eqref{eq:hamiltonian}. Two main results have been obtained: the first one has been the identification of an infinite family of two-site product integrable initial states. These are related to soliton-non-preserving boundary conditions in an appropriate transverse direction. The second one has been the characterization of the steady state reached at large times after the quench in terms of the corresponding rapidity distribution functions, which we computed analytically.

We stress that, albeit relying on some assumptions, the QTM approach outlined in this work is independent of the knowledge of the quasilocal conservation laws of the model or of the overlaps between the initial state and the eigenstates of the Hamiltonian. Conversely,  it relies on the fusion properties of the boundary transfer matrices corresponding to the integrable states. While we have used the latter in this work, their derivation is postponed to our second paper devoted to quenches in the $SU(3)$-invariant Hamiltonian \eqref{eq:hamiltonian}.

Motivated by existing results in the literature, we have conjectured analytic formulas for the overlaps between the integrable states and the eigenstates of the Hamiltonian. Based on this conjecture, we have verified that the rapidity distribution functions of the post-quench steady state obtained by the Quench Action method, coincide with those derived within the QTM formalism. These findings confirm the existence of some connection between integrable states and the possibility of deriving exact overlap formulas with the eigenstates of generic integrable Hamiltonians. Finally, as a direct physical application of our results, we have presented explicit predictions for the dynamics of the entanglement entropy after the quench.

The QTM approach applied in this work reveals several interesting directions to be explored, as it could be straightforwardly generalized to different nested models. An intriguing question pertains the application of these ideas to models defined on the continuum, including multi-component Fermi and Bose gases. These important aspects are currently under investigation.

\section*{Acknowledgments}

PC acknowledges support from ERC under Consolidator grant  number 771536 (NEMO). EV acknowledges support by the EPSRC under grant EP/N01930X. BP was supported by the BME-Nanotechnology FIKP grant of EMMI (BME FIKP-NAT), by the National Research Development and Innovation Office (NKFIH) (K-2016 grant no. 119204), and by the “Premium” Postdoctoral Program of the Hungarian Academy of Sciences. Part of this work has been carried out during a visit of BP and EV to SISSA whose hospitality is kindly acknowledged.

\appendix
\section{Proof of integrability}
\label{sec:proof_of_integrability}

In this appendix we prove that boundary states satisfy the relation \eqref{eq:starrel}, and hence are annihilated by all odd charges of the model. 

For this sake, let us introduce a pictorial representation of the different objects used in our construction. 
The three-dimensional local physical or auxilliary spaces are represented by oriented lines, whose orientation distinguishes between the fundamental (``soliton'') or conjugate (``antisoliton'') representation of $SU(3)$. The R matrices \eqref{eq:Rmatrix} and \eqref{eq:crossing} are represented as
\be 
\begin{tikzpicture}[scale=1.15,baseline={([yshift=-.5ex]current bounding box.center)}]
\draw[line width=0.5] (-0.5,-0.5) -- (0.5,0.5);
\draw[line width=0.5] (0.5,-0.5) -- (-0.5,0.5); 
\draw[-latex, line width=1.5] (-0.5,-0.5) -- (-0.35,-0.35);
\draw[-latex, line width=1.5] (0.5,-0.5) -- (0.35,-0.35);
\draw[rounded corners=2] (-0.14,-0.14) --(0,-0.2) --  (0.14,-0.14);
\node at (0,-0.4) {$\lambda$};
\node at (-0.5,-0.8) {\small $1$};
\node at (0.5,-0.8) {\small $2$};
\end{tikzpicture} =   \begin{tikzpicture}[scale=1.15,baseline={([yshift=-.5ex]current bounding box.center)}]
\draw[line width=0.5] (-0.5,-0.5) -- (0.5,0.5);
\draw[line width=0.5] (0.5,-0.5) -- (-0.5,0.5); 
\draw[-latex, line width=1.5] (-0.5,-0.5) -- (-0.65,-0.65);
\draw[-latex, line width=1.5] (0.5,-0.5) -- (0.65,-0.65);
\draw[rounded corners=2] (-0.14,-0.14) --(0,-0.2) --  (0.14,-0.14);
\node at (0,-0.4) {$\lambda$};
\node at (-0.5,-0.8) {\small $1$};
\node at (0.5,-0.8) {\small $2$};
\end{tikzpicture} = R_{12}(\lambda)\,,
\qquad 
\begin{tikzpicture}[scale=1.15,baseline={([yshift=-.5ex]current bounding box.center)}]
\draw[line width=0.5] (-0.5,-0.5) -- (0.5,0.5);
\draw[line width=0.5] (0.5,-0.5) -- (-0.5,0.5); 
\draw[-latex, line width=1.5] (-0.5,-0.5) -- (-0.35,-0.35);
\draw[-latex, line width=1.5] (0.5,-0.5) -- (0.65,-0.65);
\draw[rounded corners=2] (-0.14,-0.14) --(0,-0.2) --  (0.14,-0.14);
\node at (0,-0.4) {$\lambda$};
\node at (-0.5,-0.8) {\small $1$};
\node at (0.5,-0.8) {\small $2$};
\end{tikzpicture} =   \begin{tikzpicture}[scale=1.15,baseline={([yshift=-.5ex]current bounding box.center)}]
\draw[line width=0.5] (-0.5,-0.5) -- (0.5,0.5);
\draw[line width=0.5] (0.5,-0.5) -- (-0.5,0.5); 
\draw[-latex, line width=1.5] (-0.5,-0.5) -- (-0.65,-0.65);
\draw[-latex, line width=1.5] (0.5,-0.5) -- (0.35,-0.35);
\draw[rounded corners=2] (-0.14,-0.14) --(0,-0.2) --  (0.14,-0.14);
\node at (0,-0.4) {$\lambda$};
\node at (-0.5,-0.8) {\small $1$};
\node at (0.5,-0.8) {\small $2$};
\end{tikzpicture} =\bar{R}_{12}(\lambda)\,,
\ee 
and the direction they act on (here, upwards) is specified by the small arc at the intersection of the two lines. The crossing operation \eqref{eq:crossing} which relates the two types of R matrices can be realized as follows : 
\be 
\begin{tikzpicture}[scale=1.15,baseline={([yshift=-.5ex]current bounding box.center)}]
\draw[line width=0.5] (-0.5,-0.5) -- (0.5,0.5);
\draw[line width=0.5] (0.5,-0.5) -- (-0.5,0.5); 
\draw[-latex, line width=1.5] (-0.5,-0.5) -- (-0.35,-0.35);
\draw[-latex, line width=1.5] (0.5,-0.5) -- (0.65,-0.65);
\draw[rounded corners=2] (-0.14,-0.14) --(0,-0.2) --  (0.14,-0.14);
\node at (0,-0.4) {\small ${\lambda}$};
\end{tikzpicture}
~~~ =~~~ 
\begin{tikzpicture}[scale=1.15,baseline={([yshift=-.5ex]current bounding box.center)}]
\draw[line width=0.5] (-0.5,-0.5) -- (0.5,0.5);
\draw[line width=0.5] (0.5,-0.5) -- (-0.5,0.5); 
\draw[-latex, line width=1.5] (-0.5,-0.5) -- (-0.35,-0.35);
\draw[-latex, line width=1.5] (-0.5,0.5) -- (-0.35,0.35);
\draw[rounded corners=2] (-0.14,-0.14) --(-0.2,0) --  (-0.14,0.14);
\node at (-0.4,0) {\small $\bar{\lambda}$};
\node at (-0.65,0.65) {\small $V$};
\node at (0.65,-0.65) {\small $V$};
\end{tikzpicture}
\ee

The transfer matrices \eqref{eq:periodic_transfer_matrix} and \eqref{eq:periodic_bar_transfer_matrix} are respectively represented as 
\bea
t(\lambda) &~=~&  \frac{1}{(\lambda+i)^L}~
\begin{tikzpicture}[scale=0.9,baseline={([yshift=-.5ex]current bounding box.center)}]
\draw[line width=0.5] (-1,0) -- (8.5,0);
\foreach \x in {0,1.5,...,8} { 
	\draw[line width=0.5] (\x,-0.75)  -- (\x,0.75);
	\draw[-latex, line width=1.5] (\x,-0.75) -- (\x,-0.6);
	\draw[rounded corners=2] (\x,-0.2) --(\x-0.14,-0.14) --  (\x-0.2,0);
	\node at (\x-0.45,-0.45) {\small $\lambda$};
}
\node at (0,-1.1) {\small $1$};
\node at (1.5,-1.1) {\small $2$};
\node at (7.5,-1.1) {\small $L$};
\draw[-latex, line width=1.5] (-0.95,0) -- (-0.8,0); 
\end{tikzpicture} 
\\
\bar{t}(\lambda) &~=~&  \frac{1}{(\lambda+i)^L}~
\begin{tikzpicture}[scale=0.9,baseline={([yshift=-.5ex]current bounding box.center)}]
\draw[line width=0.5] (-1,0) -- (8.5,0);
\foreach \x in {0,1.5,...,8} { 
	\draw[line width=0.5] (\x,-0.75)  -- (\x,0.75);
	\draw[-latex, line width=1.5] (\x,-0.75) -- (\x,-0.6);
	\draw[rounded corners=2] (\x,-0.2) --(\x+0.14,-0.14) --  (\x+0.2,0);
	\node at (\x+0.45,-0.45) {\small $\lambda$};
}
\node at (0,-1.1) {\small $1$};
\node at (1.5,-1.1) {\small $2$};
\node at (7.5,-1.1) {\small $L$};
\draw[-latex, line width=1.5] (8.45,0) -- (8.3,0);
\end{tikzpicture} \,, 
\eea
where the horizontal line represents the auxilliary space $h_j$, which is understood to be traced over. 

Similar pictorial notations can be introduced as well for the various kinds of reflection matrices, an exhaustive description of which can be found in \cite{AACD04}. 
In particular we will be interested in the reflection equation \eqref{eq:reflection_1}  for the soliton-non-preserving reflection matrices, which reads pictorially 
\be
\begin{tikzpicture}[scale=1.25,baseline={([yshift=-.5ex]current bounding box.center)}]
\draw[line width=0.5,rounded corners=5] (3,1) -- (0,1) -- (0,0.5);
\draw[line width=0.5,rounded corners=5] (-1,0.5) --  (-1,1.5);
\draw[line width=0.5,] (0,0.5) arc (0:-180:0.5) ;
\draw[line width=0.5,] (2,0.5) arc (0:-180:0.5) ;
\draw[line width=0.5,rounded corners=5] (1,0.5) --  (1,1.5);
\draw[line width=0.5,rounded corners=5] (2,0.5) --  (2,1.5);
\draw[line width=0.5,rounded corners=1] (2.2,1) --(2.14,1.14) -- (2,1.2);
\draw[line width=0.5,rounded corners=1] (2.2,1) --(2.14,1.14) -- (2,1.2);
\draw[line width=0.5,rounded corners=1] (1.2,1) --(1.14,1-0.14) -- (1,0.8);
\node at (1.5,-0.25) {\footnotesize $K(\lambda)$};
\node at (-0.5,-0.25) {\footnotesize $K(\mu)$};
\node at (1.5,0.7) {\footnotesize $\lambda+\mu$};
\node at (2.5,1.3) {\footnotesize $\lambda-\mu$};
\draw[-latex, line width=1.5] (2.95,1) -- (2.8,1);
\draw[-latex, line width=1.5] (2,1.5) -- (2.,1.35);
\draw[-latex, line width=1.5] (1,1.5) -- (1.,1.35);
\end{tikzpicture} 
\quad
=
\quad
\begin{tikzpicture}[scale=1.25,baseline={([yshift=-.5ex]current bounding box.center)}]

\draw[line width=0.5,rounded corners=5]  (4,1.5) -- (4,0.5);
\draw[line width=0.5,rounded corners=5] (3,0.5) --  (3,1) -- (0.5,1);
\draw[line width=0.5,] (4,0.5) arc (0:-180:0.5) ;
\draw[line width=0.5,] (2,0.5) arc (0:-180:0.5) ;
\draw[line width=0.5,rounded corners=5] (1,0.5) --  (1,1.5);
\draw[line width=0.5,rounded corners=5] (2,0.5) --  (2,1.5);
\draw[line width=0.5,rounded corners=1] (2.2,1) --(2.14,1.14) -- (2,1.2);
\draw[line width=0.5,rounded corners=1] (2.2,1) --(2.14,1.14) -- (2,1.2);
\draw[line width=0.5,rounded corners=1] (1.2,1) --(1.14,1-0.14) -- (1,0.8);
\node at (1.5,-0.25) {\footnotesize $K(\lambda)$};
\node at (3.5,-0.25) {\footnotesize $K(\mu)$};
\node at (1.5,0.7) {\footnotesize $\lambda-\mu$};
\node at (2.5,1.3) {\footnotesize $\lambda+\mu$};
\draw[-latex, line width=1.5] (4,1.5) -- (4.,1.35);
\draw[-latex, line width=1.5] (2,1.5) -- (2.,1.35);
\draw[-latex, line width=1.5] (1,1.5) -- (1.,1.35);
\end{tikzpicture} 
\label{reflectionpic}
\ee

Setting $\lambda=-3i/4$, $\mu \to -\mu-3i/4$ and using crossing, this can be rewritten as 
\be
\begin{tikzpicture}[scale=1.25,baseline={([yshift=-.5ex]current bounding box.center)}]
\draw[line width=0.5,rounded corners=5] (3,1) -- (0,1) -- (0,0.5);
\draw[line width=0.5,rounded corners=5] (-1,0.5) --  (-1,1.5);
\draw[line width=0.5,] (0,0.5) arc (0:-180:0.5) ;
\draw[line width=0.5,] (2,0.5) arc (0:-180:0.5) ;
\draw[line width=0.5,rounded corners=5] (1,0.5) --  (1,1.5);
\draw[line width=0.5,rounded corners=5] (2,0.5) --  (2,1.5);
\draw[line width=0.5,rounded corners=1] (1.8,1) --(2-0.14,1-0.14) -- (2,0.8);
\draw[line width=0.5,rounded corners=1] (0.8,1) --(1-.14,1-0.14) -- (1,0.8);
\node at (1.5,-0.25) {\footnotesize $\tilde{K}(-3i/4)$};
\node at (-0.5,-0.25) {\footnotesize $\tilde{K}(-\mu-3i/4)$};
\node at (0.5,0.7) {\footnotesize $\mu$};
\node at (1.5,0.7) {\footnotesize $\mu$};
\node at (2,1.7) {\footnotesize $V$};
\node at (3.2,1) {\footnotesize $V$};
\end{tikzpicture} 
\quad
=
\quad
\begin{tikzpicture}[scale=1.25,baseline={([yshift=-.5ex]current bounding box.center)}]

\draw[line width=0.5,rounded corners=5]  (4,1.5) -- (4,0.5);
\draw[line width=0.5,rounded corners=5] (3,0.5) --  (3,1) -- (0.5,1);
\draw[line width=0.5,] (4,0.5) arc (0:-180:0.5) ;
\draw[line width=0.5,] (2,0.5) arc (0:-180:0.5) ;
\draw[line width=0.5,rounded corners=5] (1,0.5) --  (1,1.5);
\draw[line width=0.5,rounded corners=5] (2,0.5) --  (2,1.5);
\draw[line width=0.5,rounded corners=1] (2.2,1) --(2.14,1-0.14) -- (2,0.8);
\draw[line width=0.5,rounded corners=1] (1.2,1) --(1.14,1-0.14) -- (1,0.8);
\node at (1.5,-0.25) {\footnotesize $\tilde{K}(-3i/4)$};
\node at (3.5,-0.25) {\footnotesize $\tilde{K}(-\mu-3i/4)$};
\node at (1.5,0.7) {\footnotesize $\mu$};
\node at (2.5,0.7) {\footnotesize $\mu$}; 
\node at (2,1.7) {\footnotesize $V$};
\node at (4,1.7) {\footnotesize $V$};
\end{tikzpicture} 
\,,
\ee
where the lines are understood to be oriented as in \eqref{reflectionpic}.
We can thus apply this relation repeatedly to obtain
\bea
\begin{tikzpicture}[scale=1.15,baseline={([yshift=-.5ex]current bounding box.center)}]

\draw[line width=0.5,rounded corners=5] (4.5,1) -- (0,1) -- (0,0.5);
\draw[line width=0.5,rounded corners=5,dotted] (4.5,1) -- (5.5,1);
\draw[line width=0.5,rounded corners=5] (-1,0.5) --  (-1,1.5);
\draw[line width=0.5,] (0,0.5) arc (0:-180:0.5) ;
\foreach \x in {1,3,...,3} { 
	\node at (\x+0.5,-0.25) {\footnotesize $\tilde{K}(-3i/4)$};
	\draw[line width=0.5,] (\x+1,0.5) arc (0:-180:0.5) ;
	\draw[line width=0.5,rounded corners=5] (\x,0.5) --  (\x,1.5);
	\draw[line width=0.5,rounded corners=5] (\x+1,0.5) --  (\x+1,1.5);
	\draw[line width=0.5,rounded corners=1] (\x+.8,1) --(\x+1-0.14,1-0.14) -- (\x+1,0.8);
	\draw[line width=0.5,rounded corners=1] (\x-0.2,1) --(\x-.14,1-0.14) -- (\x,0.8);
	\node at (\x-.5,0.7) {\footnotesize $\mu$};
	\node at (\x+.5,0.7) {\footnotesize $\mu$};
	\node at (\x+1,1.7) {\footnotesize $V$};
}
\node at (-0.5,-0.25) {\footnotesize $\tilde{K}(-\mu-3i/4)$};
\node at (5.7,1) {\footnotesize $V$};
\end{tikzpicture} 
\nonumber\\
=
\begin{tikzpicture}[scale=1.15,baseline={([yshift=-.5ex]current bounding box.center)}]

\draw[line width=0.5,rounded corners=5]  (4,1.5) -- (4,0.5);
\draw[line width=0.5,rounded corners=5] (3,0.5) --  (3,1) -- (-1.5,1);
\draw[line width=0.5,rounded corners=5,dotted] (-1.5,1)--(-2.5,1);
\draw[line width=0.5,] (4,0.5) arc (0:-180:0.5) ;
\node at (3.5,-0.25) {\footnotesize $\tilde{K}(-\mu-3i/4)$};
\foreach \x in {-1,1,...,1} { 
	\node at (\x+1,1.7) {\footnotesize $V$};
	\node at (\x+0.5,-0.25) {\footnotesize $\tilde{K}(-3i/4)$};
	\draw[line width=0.5,] (\x+1,0.5) arc (0:-180:0.5) ;
	\draw[line width=0.5,rounded corners=5] (\x,0.5) --  (\x,1.5);
	\draw[line width=0.5,rounded corners=5] (\x+1,0.5) --  (\x+1,1.5);
	\node at (\x+.5,0.7) {\footnotesize $\mu$};
	\node at (\x+1.5,0.7) {\footnotesize $\mu$}; 
	\draw[line width=0.5,rounded corners=1] (\x+1.2,1) --(\x+1.14,1-0.14) -- (\x+1,0.8);
	\draw[line width=0.5,rounded corners=1] (\x+.2,1) --(\x+.14,1-0.14) -- (\x,0.8);
}

\node at (4,1.7) {\footnotesize $V$};
\end{tikzpicture} 
\eea
For $K$ invertible $\tilde{K}$ is also invertible. From here, we can then repeat step by step the derivation outlined in \cite{PiPV17_II}, and conclude
\be 
t(\mu) | \Psi_0 \rangle = \Pi t(\mu) \Pi | \Psi_0 \rangle \,.
\label{eq:to_prove}
\ee  

\section{Integrable states corresponding to non-invertible $K$-matrices}
\label{sec:non-invertible_K}

In this appendix, we show that it is possible to construct integrable states corresponding to non-invertible $K$-matrices. One possibility is to project onto one $SU(2)$ sector, for instance by taking 
\be
K^-(\lambda) = 
\left( 
\begin{array}{ccc}
	\kappa_{1,1}(\lambda) &  \kappa_{1,2}(\lambda) & 0 \\ 
	\kappa_{2,1}(\lambda) &  \kappa_{2,2}(\lambda) & 0 \\ 
	0 & 0 & 0 \\ 
\end{array}
\right) \,, 
\label{eq:Knoninvertible}
\ee
where 
\be \left( 
\begin{array}{cc}
	\kappa_{1,1}(\lambda) &  \kappa_{1,2}(\lambda)  \\ 
	\kappa_{2,1}(\lambda) &  \kappa_{2,2}(\lambda) \\ 
\end{array}\right) ,
\ee 
is a solution of the $SU(2)$ reflection equations. The latter involve the $SU(2)$ R matrix intertwining two fundamental $SU(2)$ representations spanned by $|1\rangle, |2\rangle$ on each site. In contrast with the $SU(3)$ case these representations are self-conjugate, therefore there is no distinction between soliton-preserving or soliton-non-preserving boundary conditions. The solutions of the reflection equations are well known in this case, and can be obtained for instance as a rational limit of the XXZ solutions described in \cite{PiPV17_II}. 

The states built out of the matrices \eqref{eq:Knoninvertible} are of the form $| \Psi_0 \rangle = | \psi_0 \rangle^{\otimes L/2}$, where $| \psi_0 \rangle = \alpha |1,1\rangle+\beta |1,2\rangle+\delta |2,1\rangle+\gamma |2,2\rangle$.
Let us decompose the action of the transfer matrix $t(\lambda)$ on such states as
 \be
t(\lambda)  |\Psi_0 \rangle = t_{1,2}(\lambda)  |\Psi_0 \rangle 
+ t_{3}(\lambda) |\Psi_0 \rangle \,,
\ee 
where $t_{1,2}(\lambda)$ and $t_{3}(\lambda)$ are the transfer matrices 
obtained by tracing over the subspaces of the auxiliary space spanned 
by $|1 \rangle, |2 \rangle$, or by $|3 \rangle$ respectively.
Looking first at $t_{3}(\lambda) |\Psi_0 \rangle$, it is easy to see that the only term surviving in the trace is that for which the identity term has been picked in each R matrix. Therefore $t_{3}(\lambda) |\Psi_0 \rangle$ is proportional to $ |\Psi_0 \rangle $, and we readily have  
\be 
t_{3}(\lambda) |\Psi_0 \rangle = \Pi t_{3}(\lambda) \Pi |\Psi_0 \rangle
\,.
\ee
Turning to $t_{1,2}(\lambda)  |\Psi_0 \rangle $, it is a direct consequence of $SU(2)-$integrability that 
\be 
t_{12}(\lambda) |\Psi_0 \rangle = \Pi t_{12}(\lambda) \Pi |\Psi_0 \rangle
\,,
\ee
as was proved in \cite{PiPV17_II}. Putting the two equations together therefore results in \eqref{eq:starrel}, which proves the integrability of the constructed state.

\section{Proof of the restriction to diagonal $K$-matrices}
\label{sec:restriction_K_matrices}

In this section we show that any state \eqref{eq:integrable_block} can be related to \eqref{eq:integrable_block_diagonal} through a global $SU(3)$ rotation. First, we can always perform a Schmidt decomposition to \eqref{eq:integrable_block} to obtain
\be
|\psi\rangle=\sum_{j=1}^{3}\alpha_j|a_j\rangle \otimes |b_j\rangle\,,
\label{eq:id_1}
\ee
where $\{|a_{j}\rangle\}_{j=1}^3$, $\{|b_{j}\rangle\}_{j=1}^3$ are two orthonormal bases for $\mathbb{C}^3$, while $\alpha_j\in\mathbb{R}$, $\alpha_j>0$. From \eqref{eq:integrable_block} it follows
\be
|\psi\rangle=\mathcal{P}_{1,2}|\psi\rangle=\sum_{j=1}^{3}\alpha_j|b_j\rangle \otimes |a_j\rangle\,,
\label{eq:id_2}
\ee
Consider first the case
\be
\alpha_j\neq \alpha_k\,,\quad \alpha_j\neq 0\,.
\label{eq:hypothesis}
\ee
Then, it must be $|a_j\rangle\propto |b_j\rangle$. Indeed, suppose this is not the case and that, for instance, $|a_1\rangle$ is not proportional to $|b_1\rangle$. Then, it must be either $\langle a_1|b_2\rangle\neq 0$ or $\langle a_1|b_3\rangle\neq 0$. Let us assume without loss of generality that the former is true. Multiplying then \eqref{eq:id_1} and \eqref{eq:id_2} by $|b_1\rangle \otimes |b_2\rangle$ we get
\be
\alpha_{2}\langle b_1|a_2\rangle=\alpha_{1}\langle b_2|a_1\rangle\,.
\ee
We conclude that also $\langle b_1|a_2\rangle\neq 0$, so that 
\be
\frac{\alpha_2}{\alpha_1}=\frac{\langle b_2|a_1\rangle}{\langle b_1|a_2\rangle}\,.
\label{eq:temp_1}
\ee 
Analogously, multiplying then \eqref{eq:id_1} and \eqref{eq:id_2} by $|a_1\rangle \otimes |a_2\rangle$ we get
\be
\alpha_{1}\langle a_2|b_1\rangle=\alpha_{2}\langle a_1|b_2\rangle\,,
\ee
so that 
\be
\frac{\alpha_2}{\alpha_1}=\frac{\langle a_2|b_1\rangle}{\langle a_1|b_2\rangle}\,.
\label{eq:temp_2}
\ee 
Using \eqref{eq:temp_1} and \eqref{eq:temp_2} we finally obtain
\be
\frac{|\alpha_2|^2}{|\alpha_1|^2}=1\,,
\ee
namely $\alpha_1=\alpha_2$. This contradicts the hypothesis \eqref{eq:hypothesis}. Then, by reductio ad absurdum we showed that in the case \eqref{eq:hypothesis} one has
\be
|\psi\rangle=\sum_{j=1}^{3}\alpha_j|a_j\rangle \otimes |a_j\rangle\,.
\label{eq:id_3}
\ee
It is evident that this state can always be rotated to \eqref{eq:integrable_block_diagonal} with the condition \eqref{eq:maximal_magnetization}.

Let us now consider the more complicated case where, for instance, $\alpha_1=\alpha_2$
\be
|\psi\rangle=\alpha(|a_1\rangle \otimes |b_1\rangle+|a_2\rangle \otimes |b_2\rangle)+\beta|a_3\rangle \otimes |b_3\rangle\,.
\label{eq:id_4}
\ee
The case $\beta=\alpha$ is trivial, so we can assume $\beta\neq \alpha$. With a reasoning similar to the one outlined above, one can prove that the Hilbert space generated by $|a_1\rangle$ and $|a_2\rangle$ coincides with the one generated by $|b_1\rangle$ and $|b_2\rangle$, and that $|a_3\rangle=\tilde{\beta} |b_3\rangle$, with $\tilde{\beta}\neq 0$. Then, we can parametrize
\bea
|b_1\rangle &=&v_{11}|a_1\rangle+v_{12}|a_2\rangle\,,\\
|b_2\rangle &=&v_{21}|a_1\rangle+v_{22}|a_2\rangle\,.
\eea
From \eqref{eq:id_2} it follows $v_{12}=v_{21}=v$. Plugging this into \eqref{eq:id_4} we obtain
\bea
|\psi\rangle=\alpha\left[v_{11}|a_1\rangle \otimes |a_1\rangle+v\left(|a_1\rangle \otimes |a_2\rangle+|a_2\rangle \otimes |a_1\rangle\right)\right.\nonumber\\
\left. + v_{22}|a_1\rangle \otimes |a_1\rangle\right]+\tilde{\beta}\beta|a_3\rangle \otimes |a_3\rangle\,.
\label{eq:id_5}
\eea
It is now easy to show that one can always perform an change of (orthonormal) basis
\bea
|c_1\rangle&=&x_{11}|a_1\rangle+x_{12}|a_2\rangle\,,\\
|c_2\rangle&=&x_{21}|a_1\rangle+x_{22}|a_2\rangle\,,
\eea
in the space generated by $|a_1\rangle$, $|a_2\rangle$ so that \eqref{eq:id_5} is cast into the diagonal form \eqref{eq:integrable_block_diagonal}.

\section{Computation of the generating function}
\label{sec:generating_function}

In this appendix we sketch for completeness the computation of the generating function \eqref{eq:def_generating_function}, following the prescription of \cite{FaEs13, FCEC14}. First, we rewrite the transfer matrix \eqref{eq:periodic_transfer_matrix} as
\be
t(\lambda)=\sum_{\{a_j\}\{b_j\}}{\rm tr}\{M^{a_L,b_L}(\lambda)\ldots M^{a_1,b_1}(\lambda)\}E_{a_L}^{b_L}\otimes\ldots\otimes E_{a_1}^{b_1}\,,
\label{eq:representation_t}
\ee
where $M^{a,b}(\lambda)$ are $3\times 3$ matrices defined by
\be
M^{(a,b)}(\lambda)=\frac{\lambda}{\lambda+i}\delta_{a,b}\mathbf{1}+\frac{i}{\lambda+i}E_{b}^{a}\,,
\ee
and $E_{b}^{a}$ is defined in \eqref{eq:eij_operator}. Using \eqref{eq:representation_t} it is straightforward to compute
\be
\fl t^{\dagger}(\lambda)t(\mu)=\sum_{\{b_j\},\{d_j\}}{\rm tr}\left\{N^{a_L,b_L}(\lambda)\ldots N^{a_1,b_1}(\lambda)\right\}E_{d_L}^{b_L}\otimes\ldots \otimes E_{d_1}^{b_1} \,,
\ee
where
\be
N^{d,b}=\sum_{a=1}^{3}\overline{M}^{a,d}(\lambda)\otimes M^{a,b}(\mu)\,.
\ee
Finally, defining
\be
\alpha[d_2,d_1,b_2,b_1]=\langle\psi|E_{d_2}^{b_2}\otimes E_{d_1}^{b_1}|\psi\rangle\,,
\ee
where $|\psi\rangle$ is given in \eqref{eq:integrable_block_diagonal}, we simply have
\be
\langle\Psi_0|t^{\dagger}(\lambda)t(\mu)|\Psi_0\rangle={\rm tr}\left[U(\lambda,\mu)^{L/2}\right]\,,
\ee
where
\be
U(\lambda,\mu)=\sum_{d_1,d_2,b_1,b_2}N^{d_2,b_2}N^{d_1,b_1}\,.
\ee
One can now verify that the leading eigenvalue of $U(\lambda,\lambda)$ is $\Lambda_0(\lambda,\lambda)=1$. Accordingly, using \eqref{eq:computation_omega}, one has
\be
\Omega_0(\lambda)=\frac{i}{2}\frac{\partial}{\partial \mu}\Lambda(\lambda,\mu)\Big|_{\mu=\lambda}\,,
\ee
where $\Lambda_0(\lambda,\mu)$ is the eigenvalue which is leading in a neighborhood of $\mu=\lambda$. Making finally use of Jacobi's formula, we obtain the final result
\be
\Omega_0(\lambda)=\frac{i}{2}\frac{{\rm tr}\left[{\rm adj}\left(U(\lambda,\lambda)-\mathbf{1}\right)\partial_{\mu}U(\lambda,\mu)\right]}{{\rm tr}\left[{\rm adj}\left(U(\lambda,\lambda)-\mathbf{1}\right)\right]}\,,
\label{eq:final_result_generating}
\ee
where ${\rm adj}[M]$ denotes the adjugate, namely the transpose of the
matrix of cofactors of the matrix $M$. Formula \eqref{eq:final_result_generating} can be easily evaluated for any choice of $\kappa_{jj}$. 
The ensuing analytic expression is however rather lengthy and will not be reported here. 


\end{document}